\documentclass{aa}
\usepackage{graphicx}
\begin{document}
%
%$   \thesaurus{11     % A&A Section 6: Form. struct. and evolut. of stars
%               (11.03.4; % ] {\bf Galaxies: individual:} $\ldots$
%                11.19.6; % ] Galaxies: structure
%                13.18.1; % ] Radio continuum: galaxies
%                11.10.1)} % ] Galaxies: jets  \item[11.01.2] Galaxies: active
%               % 03.20.2)} % ] Techniques: interferometric
   \title{A relativistic helical jet in the $\gamma $-ray AGN 1156+295}

%   \subtitle{I. Overviewing the $\kappa$-mechanism}
   \author{X.~Y.~Hong           \inst{1,2} \and D.~R.~Jiang   \inst{1,2}
           \and L.~I.~Gurvits   \inst{3}   \and M.~A.~Garrett \inst{3}   \and S.~T.~Garrington \inst{4}
           \and R.~T.~Schilizzi \inst{5,6}   \and R.~D.~Nan     \inst{2} \and H.~Hirabayashi   \inst{7}
           \and W.~H.~Wang    \inst{1,2} \and G.~D.~Nicolson   \inst{8}}

   \institute{Shanghai Astronomical Observatory, Chinese Academy of
              Sciences, 80 Nandan Road, Shanghai 200030, China
  \and
              National Astronomical Observatories, Chinese Academy of
              Sciences, Beijing 100012, China
  \and
              Joint Institute for VLBI in Europe, Postbus 2,
              7990~AA Dwingeloo, The Netherlands
  \and
              Jodrell Bank Observatory, University of Manchester,
              Macclesfield, Cheshire SK11--9DL, UK
   \and
              ASTRON, Postbus 2,
              7990~AA Dwingeloo, The Netherlands
    \and
              Leiden Observatory, PO Box 9513, 2300, RA Leiden,
              The Netherlands
  \and
              Institute of Space and Astronautical Science,
              3-1-1 Yoshinodai, Sagamihara, Kanagawa 229-8510, Japan
  \and
              Hartebeesthoek Radio Astronomy Observatory, Krugersdorp 1740, South Africa
%            \thanks{ }
             }

  \offprints{X.~Y.~Hong, \\ \email{xhong@center.shao.ac.cn}}

  \date{Received 31 January 2003/ accepted 19 November 2003}
%\authorrunning
%\titlerrunning

  \abstract{ We present the results of  a number of high resolution radio
observations of  the AGN 1156+295. These include multi-epoch and
multi-frequency VLBI, VSOP, MERLIN and VLA observations made over
a period of 50 months. The 5~GHz MERLIN images trace a straight
jet extending to $\sim 2\arcsec$ at ${\rm P.A.} \sim -18\degr$.
Extended low brightness emission was detected in the MERLIN
observation at 1.6~GHz and the VLA observation at 8.5~GHz with a
bend of $\sim 90\degr$ at the end of the 2 arcsecond jet. A region
of similar diffuse emission is also seen about 2 arcseconds south
of the radio core. The VLBI images of the blazar reveal a core-jet
structure with an oscillating jet  on a milli-arcsecond (mas)
scale which aligns with the arcsecond jet at a distance of several
tens of milli-arcseconds from the core.  This probably indicates
that the orientation of the jet structure is close to the line of
sight, with the northern jet being relativistically beamed toward
us. In this scenario the diffuse emission to the north and south
is not beamed and appears symmetrical. For the northern jet at the
mas scale, proper motions of $13.7\pm3.5, 10.6\pm2.8$, and
$11.8\pm2.8\;c$ are measured in three distinct components of the
jet ($q_0=0.5$, $H_0=65~$km s$^{-1}$Mpc$^{-1}$ are used through
out this paper). Highly polarised emission is detected on VLBI
scales in the region in which the jet bends sharply to the
north-west. The spectral index distribution of the source shows
that the strongest compact component has a flat spectrum, and the
extended jet has a steep spectrum. A helical trajectory along the
surface of a cone was proposed based on the conservation laws for
kinetic energy and momentum to explain the observed phenomena,
which is in a good agreement with the observed results  on scales
of 1 mas to 1 arcsec.
     }

\maketitle

%________________________________________________________________

\section{Introduction}

The AGN 1156+295 is extremely variable over a broad range of the
electromagnetic spectrum, from radio waves to $\gamma$-rays. At a
redshift of $ z=0.729$ (\cite{VCV}), it has been classified as
both a Highly Polarised Quasar (HPQ) and an Optically Violent
Variable (OVV) source (\cite{W1}, 1992, \cite{G1}). In its active
phase, the source shows fluctuations at optical wavelengths with
an amplitude of $\sim 5 - 7 \% $ on a time scale of 0.5 hour
(\cite{W1}). There are also large and rapid changes of the optical
linear polarization, with the percentage of polarised flux density
changing between $1-29\%$ (\cite{W2}). It is one of the class of
radio sources detected by EGRET (Energetic Gamma Ray Experiment
Telescope) on the Compton Gamma Ray Observatory. While quiescent
$\gamma$-ray emission remains undetected in the source, three
strong flares ($\sim 3.9\times10^{-10}$~Jy) were detected by EGRET
at energies $> 100$~MeV during the period from 1992 to 1996
(\cite{TH}, \cite{M3}, \cite{H2}).

%__________________________________________________ One column table
   \begin{table*}
   \centering
      \caption[]{The epochs, frequencies, and arrays of the observations described in this paper}
         \label{ObsSta}
       \[
     \begin{tabular}{rlrcclll}
     \hline \hline
Obs.     &    Epoch  &   Band   & Bandwidth & On-source & Array  &Telescopes$^{a}$               & Correlator  \\
No.      &           &   (GHz)  &   (MHz)   &  (hrs)    &        &                               &\\\hline
 1       &   1996.43 &  5.0     &   64      &   0.06    & VLBA   & All 10                        &  VLBA$^{b}$ \\
 2       &   1997.14 &  5.0     &   28      &   10      & EVN +  & Ef Sh Cm Jb Mc On Hh Ur Wb Tr &  MPIfR$^{c}$\\
         &           &  5.0     &   14      &   10      & MERLIN &                               &  MERLIN \\
 3       &   1997.41 &  1.6     &   14      &   10      & MERLIN &                               &  MERLIN \\
 4       &   1997.42 &  1.6     &   32      &   2.7     & VSOP   & HALCA + VLBA (all 10)         &  VLBA   \\
 5       &   1998.12 &  5.0     &   32      &   2.5     & Global & VLBA Ef Sh Jb Mc Nt On Wb Tr  &  VLBA   \\
 6       &   1999.01 &  15.0    &   64      &   0.5     & VLBA   & All 10                        &  VLBA   \\
 7       &   1999.14 &  5.0     &   28      &   10      & EVN  + & Ef Sh Cm Jb Nt On Hh Ur Wb Tr &  MPIfR  \\
         &           &  5.0     &   14      &   10      & MERLIN &                               &  MERLIN \\
 8       &   1999.45 &  5.0     &   32      &   2.5     & Global & VLBA Ef Jb Mc Nt On Wb        &  VLBA   \\
 9$^{d}$ &   2000.15 &  1.6     &   32      &   0.5     & VLBA   & All 10                        &  VLBA   \\
 10      &   2000.92 &  8.5     &   50      &   0.15    & VLA    & All 27                        &  VLA    \\
         &           &  22.5    &   50      &   0.15    & VLA    & All 27                        &  VLA    \\\hline
\end{tabular}
   \]
   \begin{list}{}{}
\item{$^{\mathrm{a}}$ Telescope codes: Ef: Effelsberg, Sh:
Shanghai, Cm: Cambridge, Jb: Jodrell Bank (MK2),  Mc: Medicina,
Nt: Noto, \\
$\;\;\;$On: Onsala,  Hh: Hartebeesthoek, Ur: Urumqi, Wb: WSRT, Tr:
Torun} \item{$^{\mathrm{b}}$ The NRAO VLBA correlator (Socorro,
USA) } \item{$^{\mathrm{c}}$ The MKIII correlator at MPIfR (Bonn,
Germany) } \item{$^{\mathrm{d}}$ Polarization observation}
\end{list}
   \end{table*}

Radio images of 1156+295 show a typical 'core~--~jet~--~lobe(s)'
morphology. On the arcsecond scale, the VLA 1.4~GHz image shows
that the source has a symmetrical structure elongated in the
north-south direction (\cite{A1}). The MERLIN image at 1.6~GHz and
VLA image at 5~GHz (\cite{M1}) show a knotty jet extending to
about $2\arcsec$ north of the core in ${\rm P.A.} = -19\degr$.
Diffuse emission is observed both to the north and south of the
core.

High resolution VLBI images of the source show a `core~--~jet'
structure with an apparently bent jet (\cite{M1}; \cite{P1};
\cite{H5}).  A wide range of superluminal velocities of the VLBI
jet components have been reported, up to $40 c$ (McHardy et al.,
1990, 1993). Piner \& Kingham (1997) reported a slower
superluminal velocity in the range of 5.4 to $~13.5~c$, based on
10 epochs of geodetic VLBI observations. Jorstad et al. (2001)
reported superluminal motion in the range of 11.8 to 18.8$~c$ on
the basis of 22 GHz VLBA (Very Long Baseline Array) observations.
Kellermann et al. (1999) and Jorstad et al. (2001) find that
$\gamma$--ray loud radio sources are more likely to exhibit
extreme superluminal motion in their radio jet components. The
data also indicate that radio flares in $\gamma$-loud sources are
much stronger than those for $\gamma$-quiet systems. In highly
variable radio sources, $\gamma$-ray outbursts are often
associated with radio flares (\cite{TL00}).

In this paper, we present the results of MERLIN observations at
1.6 and 5~GHz, VLA observations at 8.5 and 22.5~GHz, and VLBI
observations conducted with the EVN (European VLBI Network), VLBA
and VSOP (VLBI Space Observatory Programme), at frequencies of
1.6, 5, and 15~GHz. The images obtained allow us to present the
morphology of the source from parsec to kilo-parsec scales. We
discuss how the radio morphology and details of the jet structure
can be explained in the framework of the standard relativistic
beaming model (\cite{BR}).

\section{The observations and data reduction}

The epochs, frequencies, and arrays of various observations
described in this paper are summarized in Table~\ref{ObsSta}. The
overall composition of the data sets listed in Table~\ref{ObsSta}
consists of observations proposed specifically for this project
(observations \#2 \& \#7), together with additional data made
available to us by other observers (observation \#1~-~\cite{F2},
\#2~-~\cite{H5}, \#3~-~MERLIN observation of a phase calibrator,
\#4~-~\cite{H3}, \#5~-~\cite{STG1}, \#6~-~\cite{G3},
\#8~-~\cite{STG2}, \#9~\&~\#10~-~\cite{H6}).

\begin{figure*}
\vspace{250pt} \includegraphics{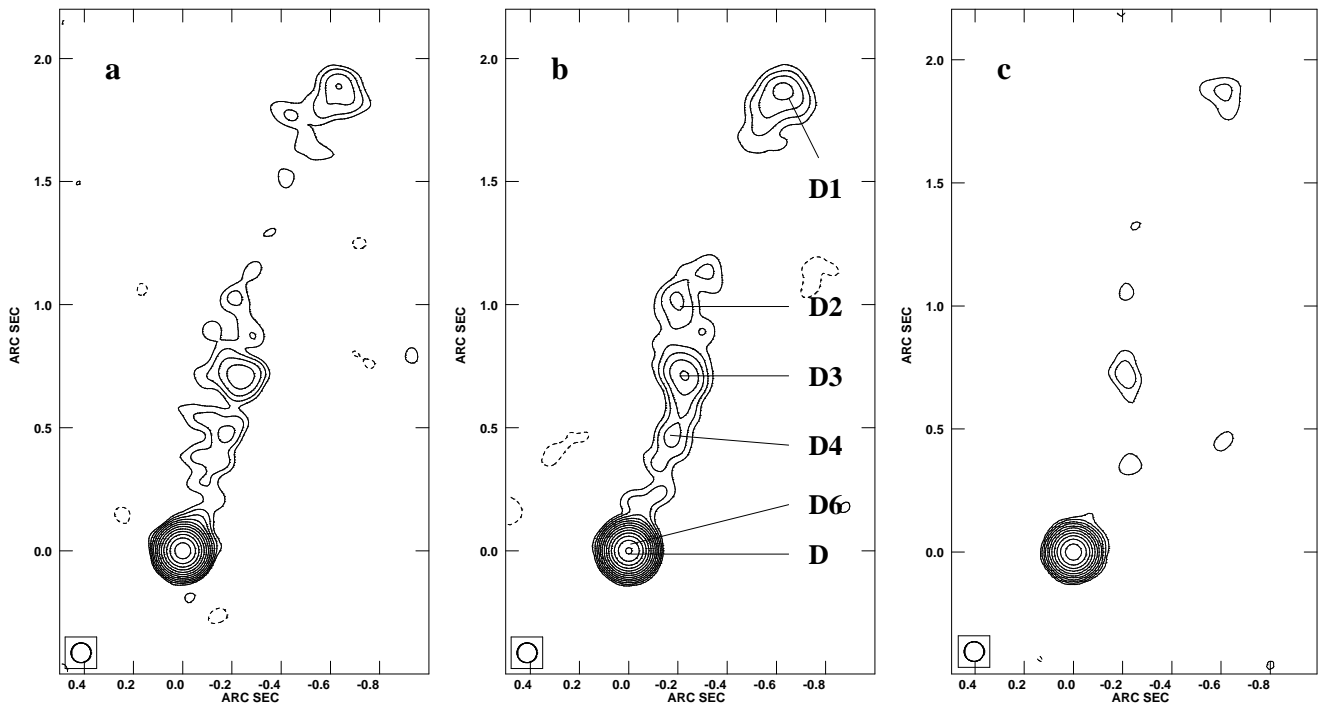} \vspace{20pt} \caption{MERLIN 5~GHz images
of 1156+295 at the epochs 1997.14 (a) and 1999.14 (b), and the
22.5~GHz VLA image at the epoch 2000.9 (c). The images were
restored with a 80~mas {\sc FWHM} Gaussian beam. Peak flux
densities are 1.6, 2.2 and 1.6~Jy/beam, the r.m.s. noise values
are 0.15, 0.15, and 0.3 mJy/beam, and the lowest contours are 0.5,
0.5 and 1.0~mJy/beam, respectively. Contour levels increase by a
factor of 2.}
\end{figure*}

\subsection{ MERLIN observations}

Two full-track observations (EVN + MERLIN) at 5 GHz were carried
out on 21 February 1997 and 19 February 1999. Another observation
(in which 1156+295 was used as a primary phase calibrator) was
made at 1.6 GHz on 28 May 1997. The MERLIN array consists of 6
antennas (Defford, Cambridge, Knockin, Darnhall, MK2, and Tabley)
at 5 GHz and 7 antennas (the same as above plus the Lovell
telescope) at 1.6 GHz (\cite{THO}).

\subsection{VLA observations}

The source was observed with the VLA in A-configuration at 8.5 and
22.5~GHz as a part of a sample of EGRET-detected AGNs in December
2000. All 27 antennas participated in the observation.

\subsection{VLBI observations}

Two epochs of full track observations with EVN (+ MERLIN), two
epochs of Global VLBI (1156+295 again was used as a phase
calibrator), and three epochs of VLBA observations of 1156+295
were carried out from 1996 to 2000 (see Table 1). The polarization
VLBA observations (epoch 2000.15) at 1.6 GHz were carried out for
24 hours for a sample of EGRET-detected AGNs.

The EVN data were correlated at the MKIII processor at MPIfR
(Bonn, Germany), the VLBA and Global VLBI data were correlated at
the NRAO (Socorro, NM, USA).

\subsection{VSOP observation}

The source 1156+295 was observed by VSOP at 1.6 GHz on 5 June 1997
as part of the VSOP in-orbit check-out procedure (\cite{H3}). In
that observation, ground support to the HALCA satellite was
provided by the Green Bank tracking station, and the co-observing
ground-based array was the VLBA.

\subsection{Data reduction}

The MERLIN data were calibrated with the suite of D-programs
(\cite{THO}). The VLA data were calibrated in AIPS (Astronomical
Image Processing System) using standard procedures and flux
density calibrators 3C48 and 3C286. Their adopted flux density
values are listed in Table~\ref{cal-flux}.

%__________________________________________________ One column table
   \begin{table}
   \centering
      \caption[]{Flux density values adopted for the VLA primary calibrators}
         \label{cal-flux}
       \[
     \begin{tabular}{c|c|c}
     \hline
Calibrator   &    X band (8.5 GHz)   &   K band (22.5 GHz)\\
             &   (Jy)      &   (Jy)   \\\hline
 3C84        &    3.22     &   1.17   \\
 3C286       &    5.18     &   2.50   \\\hline
\end{tabular}
   \]
\end{table}

The VLBI data were calibrated and corrected for residual delay and
delay rate using the standard AIPS analysis tasks. For the
polarization observations (epoch 2000.15), the solution for the
instrumental polarization (D-terms) was based on 16 scans of the
calibrator DA193, covering a large range of parallactic angles.
The absolute polarization angle of the calibrator, DA193, was
assumed to be -58{\degr} based on the VLA measurements in D
configuration at 1.67~GHz on 17 July 2000 (C. Carilli, private
communication). This was the measurement closest in time of the
absolute angle available, 0.4 years apart from the epoch of our
observation.

Post-processing including editing, phase and amplitude
self-calibration, and imaging of the data were conducted in the
AIPS and DIFMAP packages (Shepherd et al. 1994). The final images
were plotted within AIPS and the Caltech VLBI packages.

The task MODELFIT in the DIFMAP program was used to fit models of
the source structure. This consisted of fitting and optimising a
small number of elliptical Gaussian components to the MERLIN, VLA
and VLBI visibility data.

We re-imaged each pair of data-sets with the same $u,v$-ranges,
cell size, and restoring beams to produce differential images and
a spectral index distribution. Each pair of data-sets is selected
from the same correlator and as close as possible in time to
extract spectral index data and to deduce the offset.

The data used in reconstructing the spectral index distributions
were obtained non-simultaneously (except the VLA observations at
8.5 and 22.5~GHz). In principle, structural variability (proper
motion and/or changes of size or brightness of the components)
will mask the true spectral index distribution. In order to
minimize this masking effect we align images obtained at different
epochs and frequencies at the position of the core assuming its
opacity shift is small comparing to our angular resolution.

\begin{figure*}
\vspace{278pt} \includegraphics{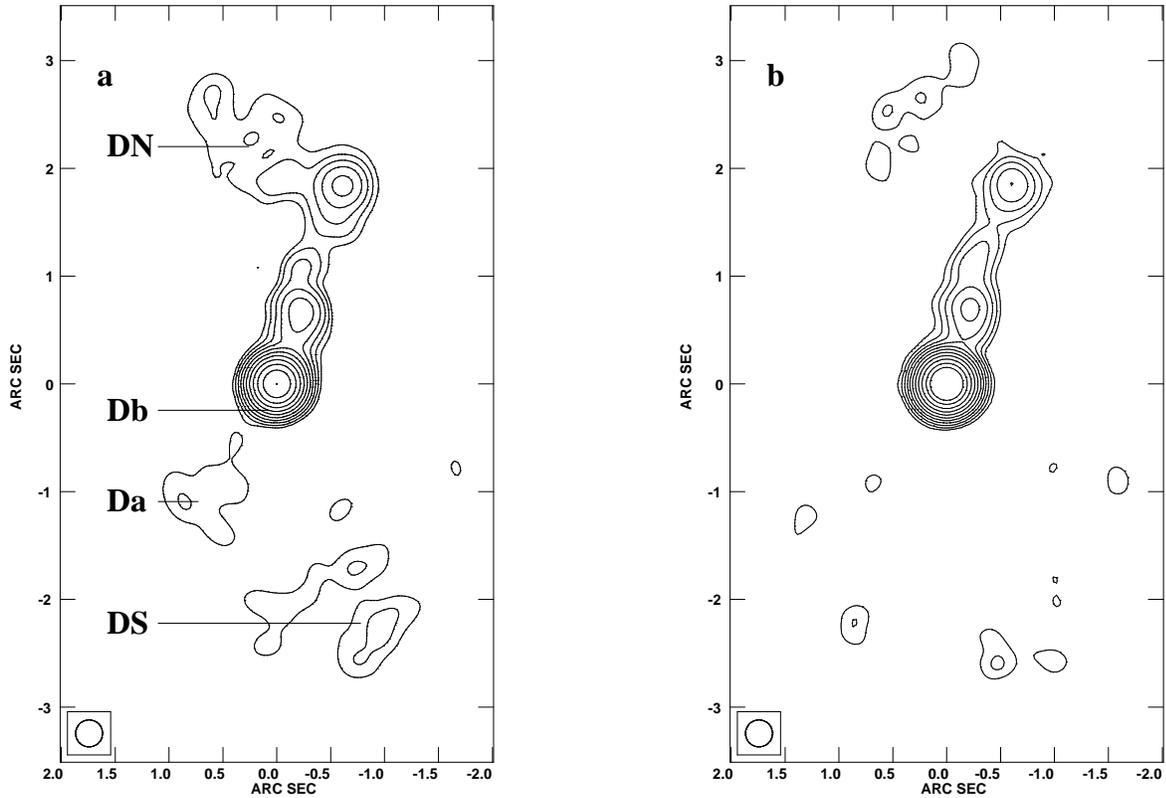} \vspace{25pt} \caption{A MERLIN 1.6~GHz
image at the epoch 1997.41 (a) and 8.5~GHz VLA image at the epoch
2000.9 (b). The images were restored with a  circular beam of 250
mas ({\sc FWHM}) in diameter. Peak flux densities are 1.55 and
1.0~Jy/beam, the r.m.s. noise levels are 0.4 and 0.13 mJy/beam and
the lowest contours are 1.5 and 0.4~mJy/beam, respectively.
Contour levels increase by a factor of 2.}
\end{figure*}

\section{Results}

\subsection{The arcsecond-scale images}

The two 5~GHz MERLIN images and the 22.5 GHz VLA image restored
with the same circular 80 mas beam are shown in Fig.~1. The peak
brightness at 5~GHz has increased by 37\% from 1.6 Jy/beam at the
epoch 1997.14 to 2.2 Jy/beam at the epoch 1999.14. This could be
related to an outburst in the radio core (see section 3.3).

The arcsecond-scale morphologies are similar with an almost
straight jet at a position angle of about $-18\degr$ with perhaps
some evidence of a sinusoidal fluctuation on the sub-arcsecond
scale (Fig. 1). A knot at $\sim0.7 \arcsec$ and a hotspot
$2\arcsec$ from the core are well detected.  Within $1 \arcsec$
from the core, the jet is resolved into several regularly spaced
knots. When the jet passes the knot, it bends slightly and ends
with a hotspot around $\sim2\arcsec$ from the core. No counter-jet
emission is detected with the MERLIN at 5~GHz and VLA at 22.5~GHz.
Only the three main discrete components (D, D3 and D1) were
detected at 22.5~GHz (Fig.~1c).

The MERLIN image at 1.6~GHz and the VLA image at 8.5~GHz are
presented in Fig.~2. We restored the two data sets with a circular
Gaussian beam of 250~mas in diameter for comparison. They both
show a $2 \arcsec$ jet at $P.A. \sim-18\degr$. Low brightness
extended emission was also detected with MERLIN at 1.6~GHz and the
VLA at 8.5~GHz. This emission is seen to bend away eastwards
($\sim 90\degr$) from the main arcsecond scale jet structure. A
region of diffuse radio emission is seen about $2\arcsec$ south of
the core too.

One explanation for the overall morphology of the source is that
it is a double-lobed radio source seen almost end-on with the
northern jet relativistically beamed towards us. Doppler boosting
makes the northern jet much brighter than its de-boosted southern
counterpart. The southern jet remains undetected except when it
becomes sub-relativistic at the end of the jet. It appears as the
diffuse emission seen to the south of the core.

Another possible explanation of the brightening of the southern
jet at distances of 1.4 and 2.2 arcseconds from the core is that
it bends closer to the line of sight, thus increasing the Doppler
boosting. However, this explanation seems to be less likely since
it would require the southern jet to be closer to the sky plane in
its inner parts while the inner northern counterpart appears to be
close to the line of sight.

\subsection{VLBI images}

VLBI images of 1156+295 at various frequencies (1.6, 5, and
15~GHz) are shown in Fig.~3 in the time sequence from (a) to (h).
The parameters of the images are summarized in
Table~\ref{parmvlbi}.

\begin{figure*}
\vspace{645pt} \includegraphics{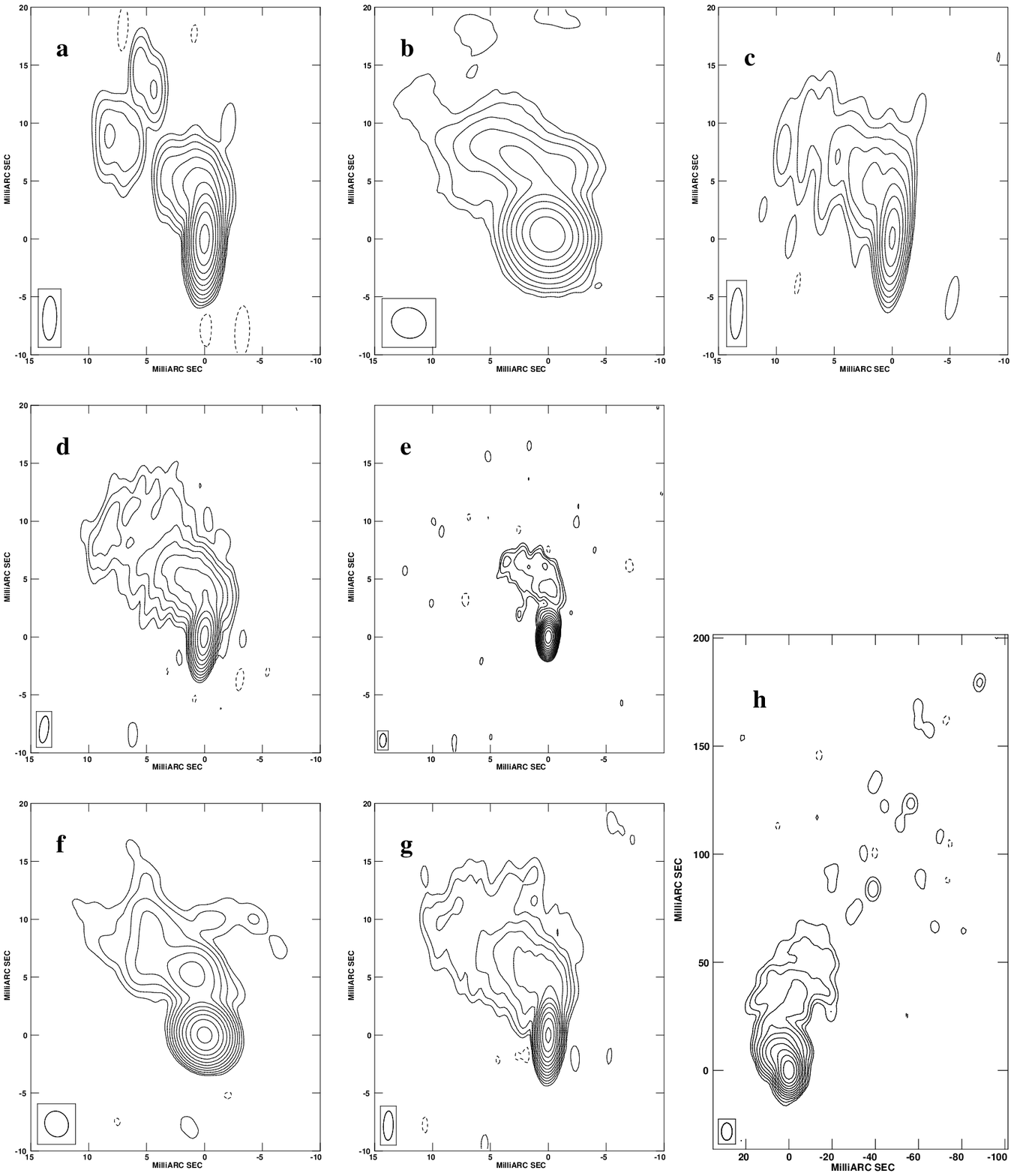} \vspace{-60pt} \caption{The VLBI Images of
1156+295. The parameters of the images are listed in
Table~\ref{parmvlbi}.
 }
\end{figure*}

%__________________________________________________ One column table
   \begin{table*}
\begin{center}
%   \centering
      \caption[]{The parameters of the images in Fig.~3}
         \label{parmvlbi}
       \[
     \begin{tabular}{llccccccc}
     \hline \hline
No.&  Epoch  & Array &Band$\;$&$S_{peak}^{*}$&$r.m.s.$& Contours                          &  Restoring Beam    \\
   &         &       & (GHz)  &(Jy/beam)&(mJy/beam  ) & (mJy/beam)                        &  Maj$\times$Min(mas), $P.A.$($\degr$)  \\\hline
a  & 1996.43 & VLBA  & 5.0    &  2.08   & $0.5\;\;$   & 1.5$\times(-1,1,2,4,...,1024)    $& $3.7\times1.2, \;\;\;\;-1.8$      \\
b  & 1997.14 & EVN   & 5.0    &  1.02   & $0.6\;\;$   & 2.0$\times(-1,1,2,4,...,\;\; 256)$& $3.0\times2.6,\;\;\;\;\;\;\;\;77$ \\
c  & 1997.42 & VSOP  & 1.6    &  0.89   & $1.0\;\;$   & 3.0$\times(-1,1,2,4,...,\;\; 256)$& $4.4\times1.15,\;\;\;\;\;-5$  \\
d  & 1998.12 &Global & 5.0    &  0.38   & 0.13        & 0.5$\times(-1,1,2,4,...,\;\;512) $& $2.3\times0.78,\;\;\;-7.7$ \\
e  & 1999.01 &VLBA   &$15.0\;$&  2.22   & $0.1\;\;$   & 0.3$\times(-1,1,2,4,...,4096)    $& $1.1\times0.63, \;\;\;-1.7$\\
f  & 1999.14 & EVN   &  5.0   &  1.34   & 0.25        & 1.0$\times(-1,1,2,4,...,1024)    $& $2.3\times2.0, \;\;\;\;\;\;\;4.2$\\
g  & 1999.45 & Global& 5.0    &  1.74   & 0.22        & 0.7$\times(-1,1,2,4,...,2048)    $& $2.5\times0.85, \;\;-1.7$\\
h  & 2000.15 &VLBA   & 1.6    &$1.0\;\;$& 0.12        & 0.5$\times(-1,1,2,4,...,1024)    $& $8.3\times5.2, \;\;\;\;-5.5$        \\
\hline
\end{tabular}
   \]
      \begin{list}{}{}
\item{$^{\mathrm{*}}$ peak brightness}
\end{list}
\end{center}
    \end{table*}

    The eight VLBI images of the source all show an oscillatory jet
    structure on mas scales. The jet initially  points
    almost exactly to the north but then bends to the north-east at 3$\sim$4 mas
    from the core. Several tens of milliarcseconds from the core it
    finally turns about $90\degr$ to the north-west, thus aligning
    with the direction of the arcsecond-scale jet.

    The five epochs of VLBI observations at 5~GHz from 1996 to 1999
    allows us to study
    the structural variations of 1156+295 in detail.
    The two global VLBI images (Fig.~3~d\&g) both with excellent
    ($u,v$)-coverage and high sensitivity, enable us to detect
    features that have relatively low surface brightness.

The highest resolution image was made with the VLBA at 15~GHz. An
additional component has been detected in the north about 1.5 mas
from the core (Fig.~3e).

The resolution of the VSOP image of 1156+295 at 1.6~GHz (Fig.~3c)
is comparable to that of ground based VLBI images at 5~GHz. This
allows us to study the source structure on the same scales but at
different frequencies. We note that a similar curved jet structure
is visible in both the 1.6~GHz VSOP and 5~GHz ground-based images.

The lower resolution VLBA image of 1156+295 at 1.6~GHz (Fig.~3h)
also displays a curved jet. In particular, the jet is observed to
turn sharply at a few tens of milliarcseconds from the core,
aligning with the direction of the kpc jet. A high degree of
linear polarization is detected in the 1.6~GHz VLBA data: two
distinct polarized components are observed in the areas where the
jet bends (see Fig.~4). A peak polarized brightness of 14.4
mJy/beam was detected in the core. The strongest polarized
component is located at 2.6 mas north of the core with a peak
polarized brightness of 18 mJy/beam. The secondary polarized
component was detected at 8.1 mas from the core at P.A.=60{\degr}
with a peak polarized brightness of 8.5 mJy/beam. The percentage
polarization of the core, the strongest component and secondary
component are 1.5\%, 2.3\%, and 7.7\%, respectively. The strongest
polarized jet component and the secondary polarized jet component
have perpendicular E-vectors to each other (Fig. 4). Highly
polarized components appear to be associated with the sharp bends
of the jet. This could be caused by a change of opacity along the
line of sight or a transition from the optically thick to the
optically thin regime.

\begin{figure}
\vspace{175pt} \includegraphics{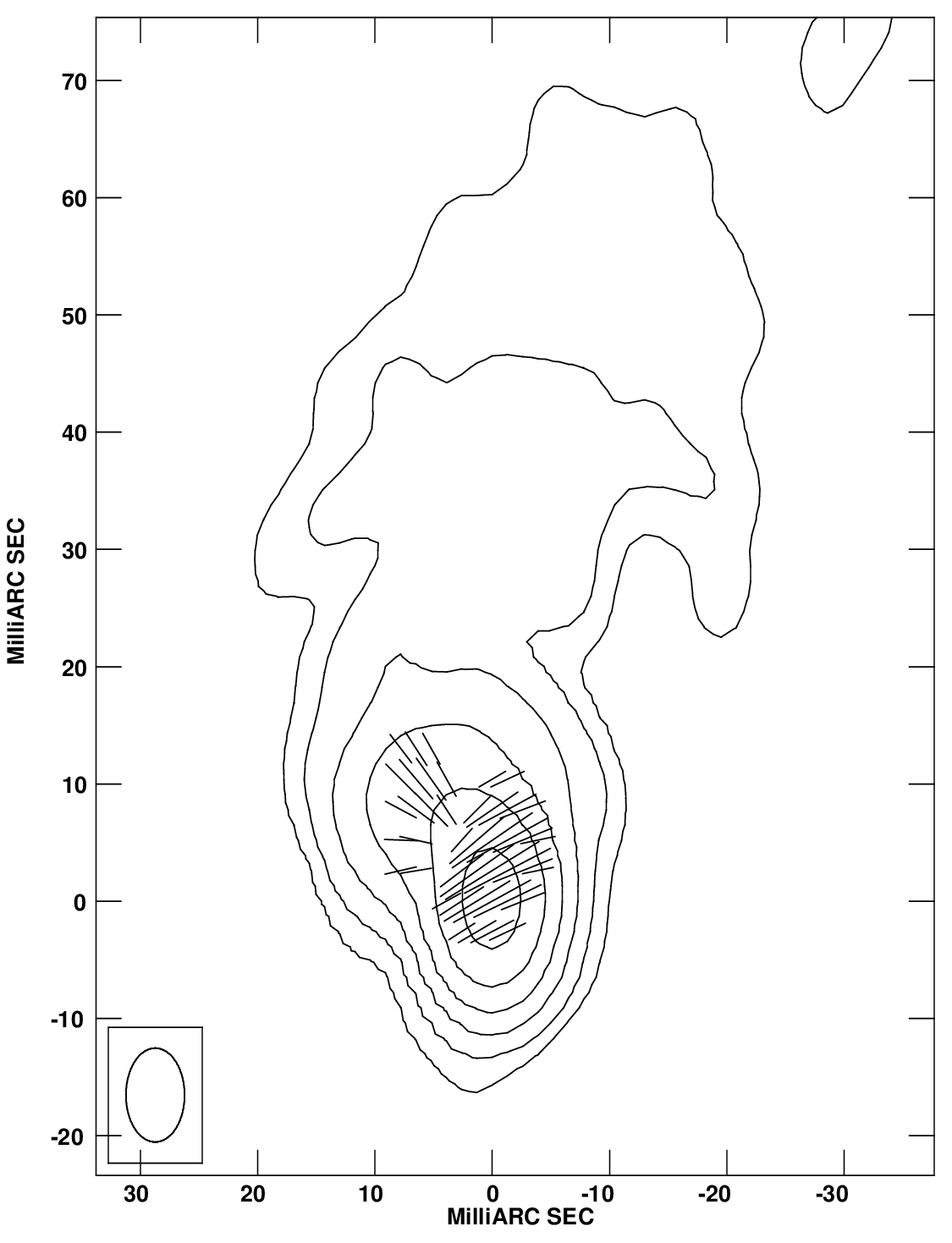} \vspace{85pt} \caption{The VLBA image at
1.6~GHz with total intensity contours at 0.5~mJy/beam $\times (-1,
1, 2, 4, 8, 16, 64, 256, 1024$), and superposed sticks show the
orientation of electric vectors (polarization line $1\arcsec$ =
1.92 Jy/beam).}
\end{figure}

The VLBI images of the source 1156+295 reveal a core-jet structure
with an oscillatory jet on mas scales. Two sharp bends in the
opposite directions occur (one curves anti-clockwise to the
north-east within a few mas of the core, while the other bends
clockwise to the north-west within a few tens of mas from the
core). The VLBI jet then aligned with the direction of the MERLIN
jet. The oscillations in the jet structure on mas scales resemble
a 3D helical pattern.

\subsection{Light Curves at Radio Frequencies}

\begin{figure*}
\vspace{410pt} \includegraphics{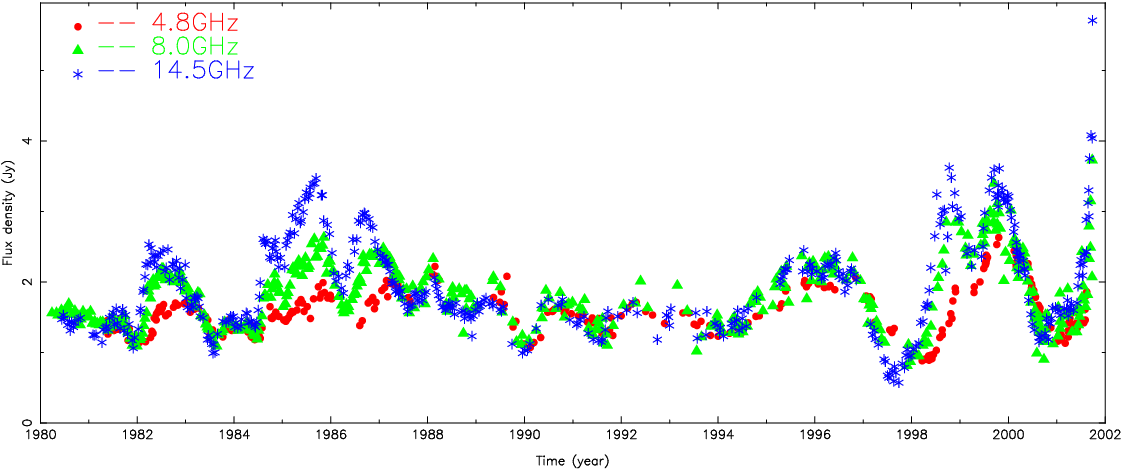} \vspace{-190pt} \caption{Light curves of
1156+295 at 4.8, 8.0 and 15~GHz from the monitoring data of the
University of Michigan Radio Astronomy Observatory.}
\end{figure*}

\begin{table*}
\centering \caption{The results of  model fitting at kiloparsec
scales} \label{model-kpc}
%  \[
\begin{tabular}{c|ccccrccc }\hline
Observation inf.&$S_{total}$&Comp& $S_{c}$ &  r      & $P.A.$  & a      & b/a &$P.A.$    \\
             & (mJy)  &          & (mJy)   & (mas)   &($\degr$)& (mas)  &     &($\degr$)  \\
   (1)       &  (2)   &(3)       & ( 4)    &  (5)    & (6)     & (7)    & (8) &  (9)   \\ \hline
1997.14      &  1668  & D        & 1608    &  0.0    & $  0.0 $&  8.4   &0.25 &15.2    \\
MERLIN       &        & D6       & 21.5    & 53.7    & $ -8.5 $& 18.6   & 1.0 & 0      \\
5 GHz        &        & D4       & 11.4    &  515    & $ -19.3$& 132    & 1.0 & 0      \\
             &        & D3       & 13.2    &  737    & $ -18.2$& 76.9   & 1.0 & 0      \\
             &        & D2       & 7.4     & 1027    & $ -10.8$& 101    & 1.0 & 0      \\
             &        & D1       & 13.9    & 1948    & $ -18.4$& 135    & 1.0 & 0      \\
{\it Errors} &        &          &{\it 8\%}&{\it9\%} & {\it 3} &{\it9\%}&     &{\it 3} \\ \hline
1999.14      & 2290   & D        & 2214    &   0.0   & $ 0.0  $&   8.6  &0.38 & 15.6   \\
 MERLIN      &        & D6       & 28.5    &  56.9   & $ -6.0 $&  23.9  & 1.0 & 0      \\
 5~GHz       &        & D4       & 13.1    &  545    & $ -20.6$&   130  & 1.0 & 0      \\
             &        & D3       & 14.0    & 743     & $ -18.0$&  70.0  & 1.0 & 0      \\
             &        & D2       & 11.2    & 1057    & $ -12.0$&   138  & 1.0 & 0      \\
             &        & D1       & 16.6    & 1938    & $ -18.3$&   128  & 1.0 & 0      \\
{\it Errors} &        &          &{\it 8\%}&{\it 9\%}& {\it 3} &{\it9\%}&     &{\it 3} \\  \hline
2000.92      &   1666 & D        & 1640    &  0.0    & $ 0.0  $&   5.8  & 0.6 & 25.1   \\
VLA          &        & D3       &  7.5    & 747     & $ -16.1$&   102  & 1   &  0     \\
22.5~GHz     &        & D1       &  4.6    &1960     & $ -17.9$&  88.0  & 1   &  0     \\
{\it Errors} &        &          &{\it 7\%}&{\it 7\%}& {\it 3} &{\it7\%}&     &{\it 3} \\ \hline
1997.41      & 1768   & D        & 1540    & 0.0     & $ 0.0  $& 10.1   & 0.7 & 15.0   \\
MERLIN       &        & D5       & 27.8    & 201     & $-27.6$ & 46.9   & 1.0 & 0      \\
1.6~GHz      &        & D4       & 15.0    & 480     & $-19.7$ & 70.1   & 1.0 & 0      \\
             &        & D3       & 42.1    & 716     & $-18.4$ &  105   & 1.0 & 0      \\
             &        & D2       & 17.4    & 1109    & $-12.8$ &  183   & 1.0 & 0      \\
             &        & D1       & 37.6    &1945     & $-18.7$ &  137   & 1.0 & 0      \\
             &        & DN       & 45.7    &2055     & $-2.0 $ & 1409   & 0.5 &59.5    \\
             &        & Db       & 20.6    & 138     & $147.6$ &  156   & 1.0 & 0      \\
             &        & Da       & 11.1    &1398     & $144.9$ &  694   & 1.0 & 0      \\
             &        & DS       & 27.6    &2152     &$-161.5$ & 1217   & 0.6 &$-75$   \\
{\it Errors} &        &          &{\it 8\%}&{\it 9\%}&{\it 3}  &{\it9\%}&     &{\it 3} \\ \hline
2000.92      & 1042   & D        & 1000    & 0.0     & $ 0.0  $& 13.5   & 0.7 &22.3    \\
VLA          &        & D5       & 2.2     & 198     &$ -26.5$ & 70.4   & 1.0 & 0      \\
8.5~GHz      &        & D4       & 3.2     & 507     &$ -22.9$ & 85.0   & 1.0 & 0      \\
             &        & D3       & 10.1    & 734     &$ -17.6$ & 135    & 1.0 & 0      \\
             &        & D2       & 3.8     & 1128    &$ -13.6$ & 231    & 1.0 & 0      \\
             &        & D1       & 9.2     & 1945    &$ -18.2$ & 135    & 1.0 & 0      \\
             &        & DN       & 6.6     & 2121    &$ -2.3 $ & 1305   & 1.0 & 0      \\
{\it Errors} &        &          &{\it 7\%}&{\it 7\%}&{\it 3}  &{\it7\%}&     &{\it 3} \\ \hline
\end{tabular}
%   \]
\end{table*}

Fig.~5 shows the light curves of 1156+295 at 4.8, 8.0, and 15~GHz
measured at the University of Michigan Radio Astronomy Observatory
(UMRAO). Several flares from 1980 to 2002 were detected.

The intensity of a flare is stronger at higher frequencies. Flares
at different frequencies do not occur simultaneously: high
frequency flares appear early. Similar results have been reported
in some other sources (\cite{B1}, \cite{WW}, \cite{zhou}). The
lags are consistent with the fact the flares components are
optically thick and short wavelengths peak first, which can be
seen directly from the light curves (Fig.~5).

The strongest flare recorded over the last 20 years started at the
beginning of 2001. The available data show a maximum in the flux
density of 5.7~Jy at 14.5~GHz in September 2001. The next
strongest flare began in October 1997 at 8.4 and 14.5~GHz, and
reached its peak in October 1998 with a flux density of 3.6~Jy at
14,.5~GHz. Another similar peak occurred in September 1999. Such
double peaked flares are also seen in the period from 1985 to
1987. The separations of the double peaks are about one year.

\begin{figure} \vspace{20pt}
%\vspace{250pt}
\includegraphics{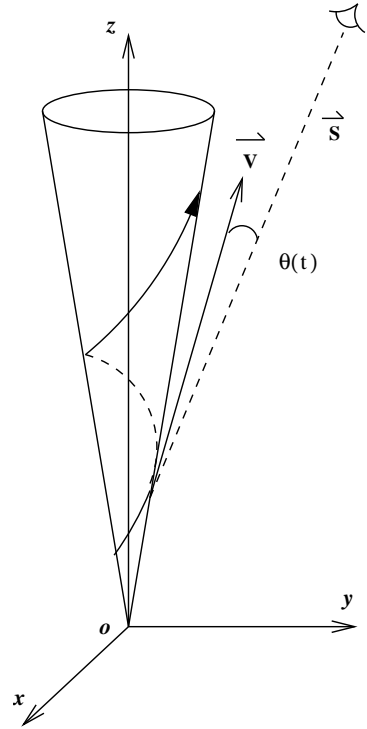} \vspace{250pt} \caption{A helical pattern for
1156+295: vector S is the direction of the line of the sight,
vector V is the direction of the velocity of the jet, $\theta(t)$
is the viewing angle between V and S.}
\end{figure}

\begin{figure}
\vspace{175pt} \includegraphics{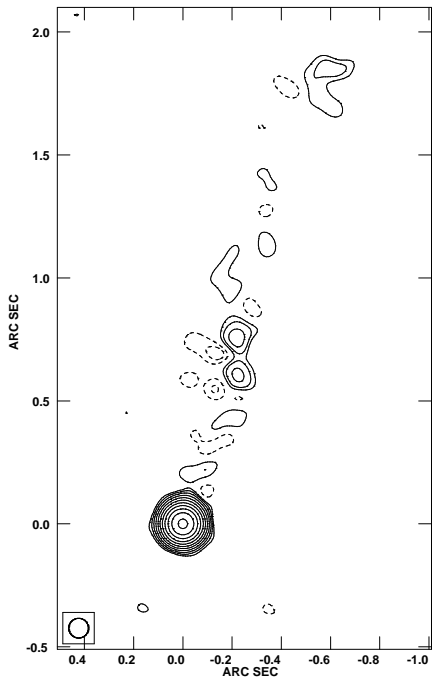} \vspace{85pt} \caption{The differential
image of the two 5~GHz MERLIN data-sets from the epoch 1999.14 and
1997.14, Peak flux density is 0.6~Jy/beam. Contour levels are
0.5~mJy/beam $\times (-4, -2, -1, 1, 2, 4, 8, 16, 32, ...$), FWHM:
80 $\times$ 80 mas.}
\end{figure}

A possible explanation for the 'double-peak'  is that only one jet
component is ejected with a helical trajectory on the surface of a
cone. The emission of the jet component is then enhanced along the
path but only when it points toward the observer. If two nearby
pitch angles (or multi pitch angles) are close to the line of the
sight, we can observe double peaked flares (or multi peaked
flares) with the maximum Doppler boosting from one ejected
component. In the case of 1156+295, this results in two distinct
peaks in the total flux density measurements. If the jet
components always travel along a helical path aligned with the
line-of-sight then double peaks should always appear in the light
curves. In particular, double peaks should appear at the outburst
near 2002. In this case, as shown in Fig. 6, we find one maximum
for Doppler boosting at the each helical cycle. Then the peak
separation time of double peaks $\Delta t_{pk}$ should be the
period of the first detectable helical orbit and can be directly
inferred as the period of the precession of the jet-base. The pure
geometrical model described here is similar to the lighthouse
model proposed by Camenzind \& Krockenberger (\cite{CK92}).
However, in the case of 1156+295, the observed period of the peak
of its flux density is somewhat longer than that in the lighthouse
model.

Two jet components ejected one after the other can also explain
the 'double-peak' pattern. If these two jet components move out
from the radio core, they might be resolved from each other by
high resolution VLBI images. Further multi-epoch VLBA observations
at higher frequencies (2cm, 7 and 3 mm) are currently underway,
which will help to clarify the issue.

The peak brightness detected in our multi-epoch VLBI observations
varies in  phase with the total flux density measurement made at
UMRAO. This indicates the most outbursts take place in the core
area and/or in the unresolved inner-jet.

\section{Morphology and structural variability on kpc-scales}

\subsection{Model fitting the MERLIN and VLA data}

The MERLIN and VLA visibility data were fitted with elliptical
Gaussian components using DIFMAP. The results of model-fitting are
presented in Table~\ref{model-kpc}. Column 1 gives the observation
epoch, array, and observed frequency. The total CLEANed flux
density of the image is listed in column 2, followed by the
component's name in column 3.  Column 4 shows the flux density of
the component. This is followed in columns 5 and 6 by the radial
distance and position angle of the component (relative to the core
component). The next three entries (columns 7 to 9) are the major
axis, axis ratio and position angle of the major axis of the
fitted component. The uncertainties of the model fitting listed in
Table~\ref{model-kpc} are estimated using the formulae given by
Fomalont (\cite{Fomalont}).

Besides the core component D, six components (D1 to D6) are fitted
in the northern jet (see Fig.~1b). Two counter jet components (Da
and Db) and two lobe components (DN and DS) are detected in the
1.6~GHz MERLIN data (as labelled in Fig.~2a). D5 is too faint to
be well fitted in the 5~GHz MERLIN data (Fig.~1a\&b). D6 is not
fitted with the 1.6~GHz MERLIN and 8.5~GHz VLA data because of the
limitation of the resolution (Fig.~2). Only three main components
(D, D1 and D3) are detected in the 22.5~GHz VLA data (Fig.~1c).

From columns 2 and 4 of Table~\ref{model-kpc}, it is clear that
more than 95~$\%$ of the flux density came from the core at all
frequencies, except 87$\%$ at 1.6~GHz. This is consistent with the
flux density of the core being enhanced by beaming effects while
the extended emission is not.

\begin{figure*}
\vspace{195pt}
 \includegraphics{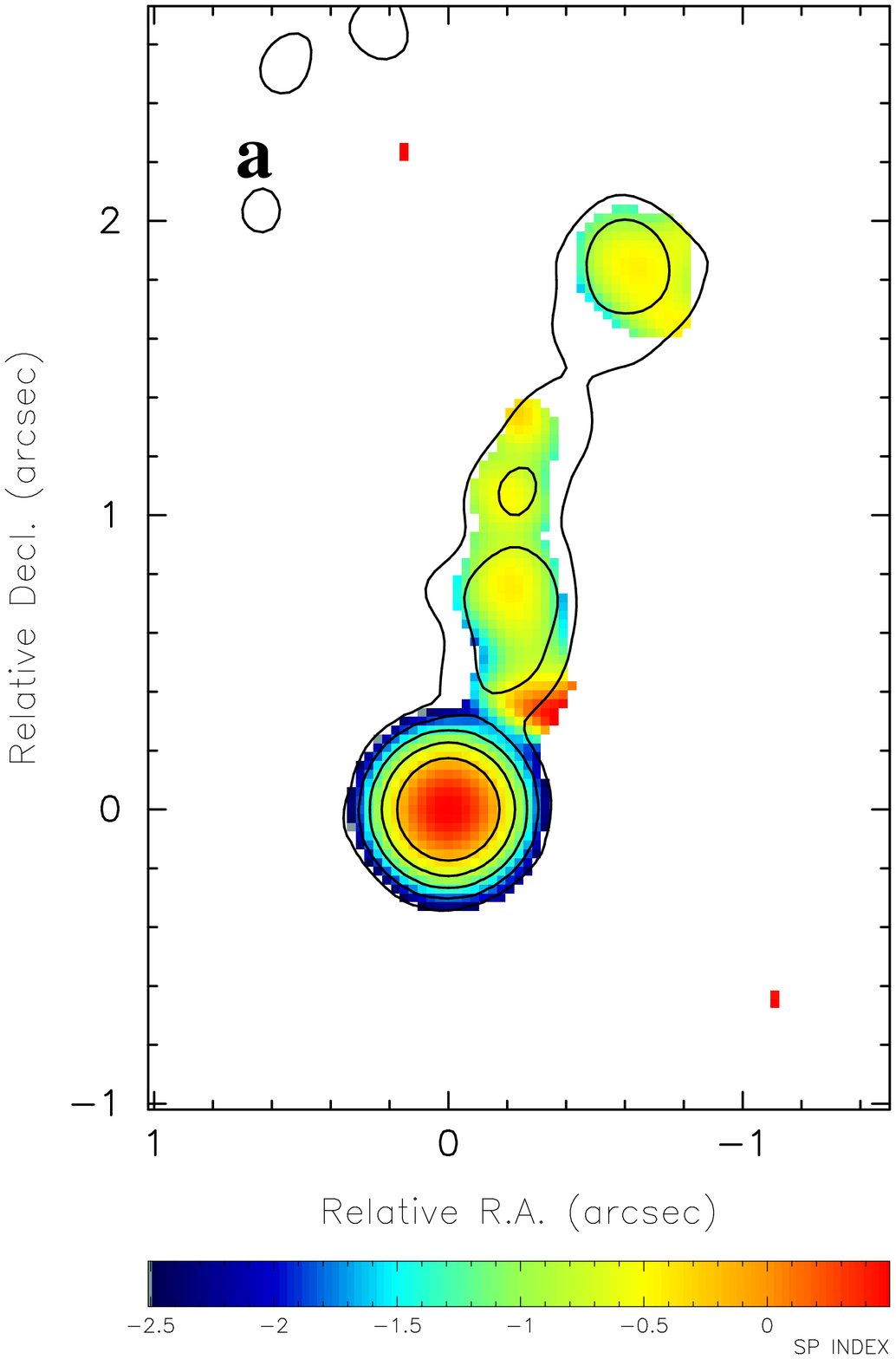} \includegraphics{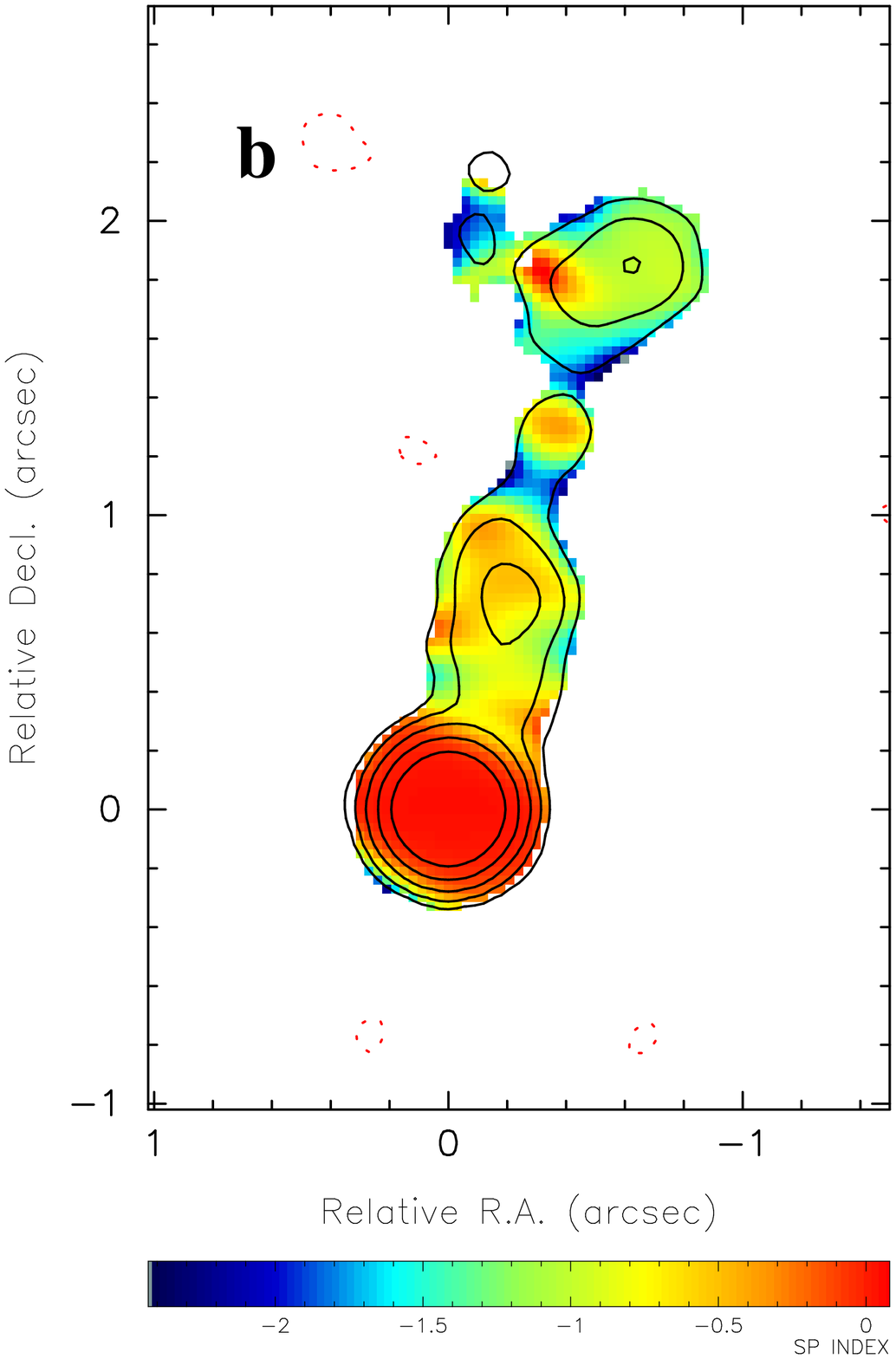} \vspace{80pt} \caption{(a): A
simultaneous spectral index distribution between 8.5 and 22.5~GHz
at the epoch 2000.92 (color), superimposed on the 8.5~GHz VLA
image; the wedge at the bottom shows the spectral index range
$-2.5$ to 0.5; (b): A 1.6 - 5~GHz non-simultaneous spectral index
distribution (color), superimposed on the 5~GHz MERLIN image at
the epoch 1997.14; the wedge at the bottom shows the spectral
index range $-2.4$ to 0.08. The contours levels in both b\&c are
0.5 mJy/beam $\times$ ($-1, 1, 4, 16, 64, 256$) with FWHM of
200$\times$200 mas.}
\end{figure*}

\begin{figure*}
\vspace{190pt}
 \includegraphics{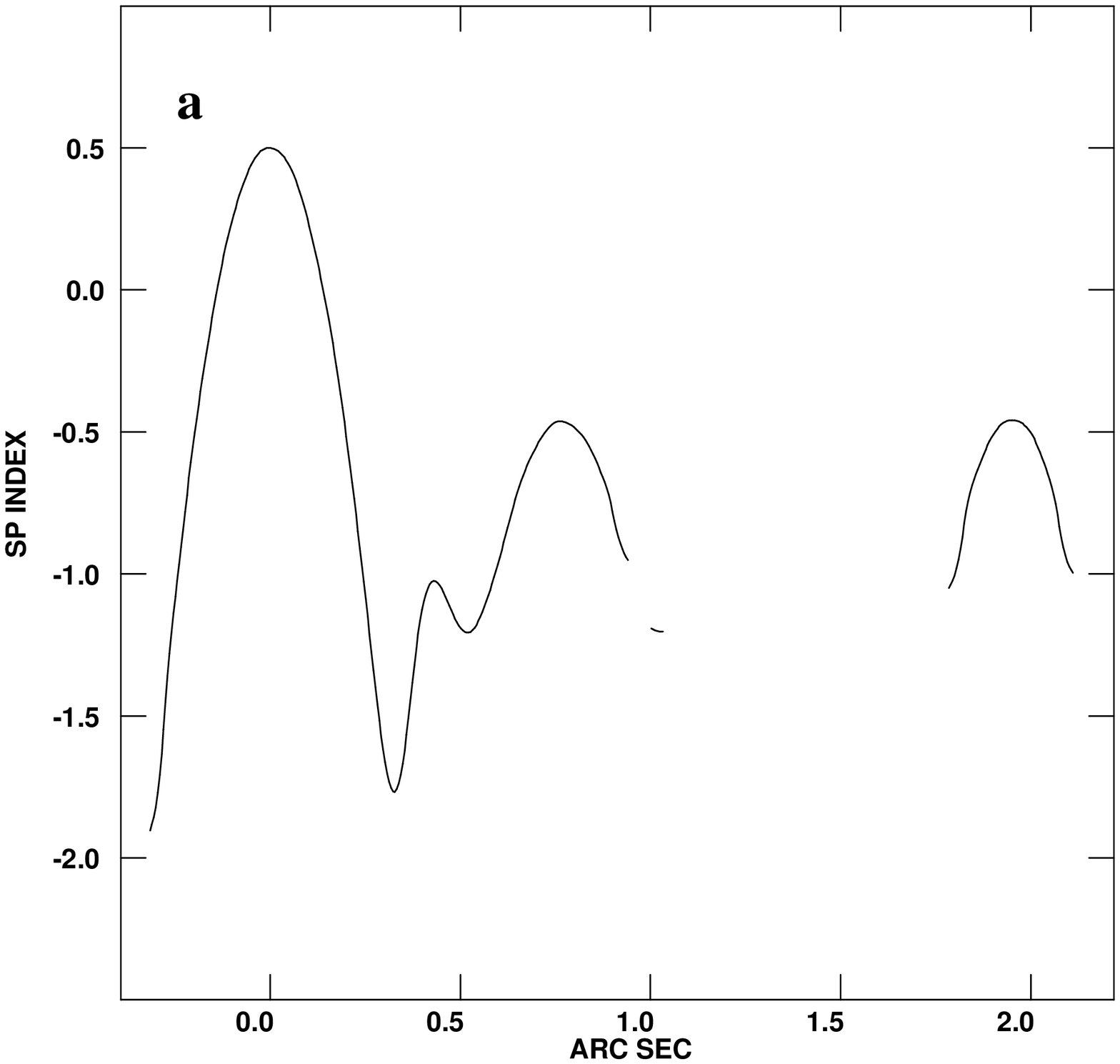} \includegraphics{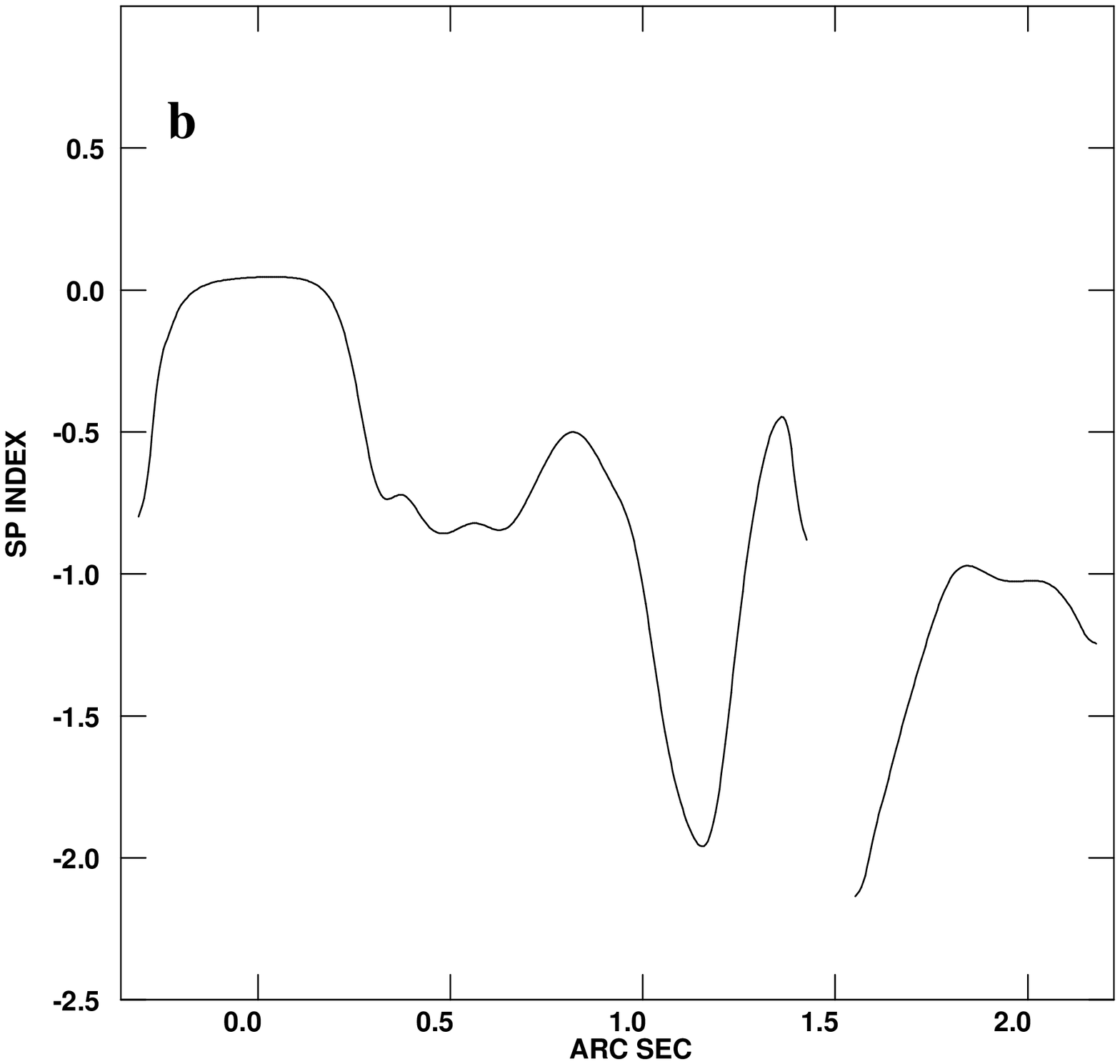} \vspace{50pt} \caption{Slice
plots of spectral index at p.a. -18{\degr}, corresponding to Fig.
8a and Fig. 8b respectively.}
\end{figure*}

The two main kpc-scale components (D1 and D3) are almost at the
same position angle, $\sim -18\degr$, in all MERLIN and VLA
images, and their FWHMs are comparable at 1.6, 8.5, and 22~GHz,
while D3 is more compact than D1 at 5~GHz. We note that D2 and D4
sometimes appear to have larger FWHMs than that of D1 and D3,
which may be attributed to using circular Gaussian components for
model fitting.

A counter jet component (Db) is fitted in the 1.6~GHz MERLIN data
(see Fig.~2a). It is located at about 138 mas from the core at
$p.a. \sim 147\degr$. Two other low brightness components are also
detected (labelled as Da and DS in Table~\ref{model-kpc} and
Fig.~2a).

The lobe to counter-lobe (the components DN and NS) flux density
ratio at a distance of 2 arcseconds from the core is
$J_{lobe}=1.66$. The distance and size of the counter-jet
component Da allow us to assume that its corresponding structural
pattern in the jet should include the components D2 and D3. Under
this assumption, the flux density ratio (D2+D3) to counter-jet Da
is $J_{kpc-jet}=5.36$. No clear jet component corresponding to Db
is evident in Fig.2a. This might be due to the corresponding jet
component being embedded in the core.

\subsection{Structural variability on kpc-scales}

The two 5~GHz MERLIN observations with a time separation of two
years (the epochs 1997.14 and 1999.14) allow us to study
structural variability. As we mentioned in section 3.3, the source
was in its lowest state around 1998, immediately before a strong
flare occurred (see Fig.~5). The model fitting results show that
the flux density of the core component D increased by 37\% from
1997 to 1999 (Table~\ref{model-kpc}), which is most easily
explained in terms of an outburst occurring in the inner core.

To study the variation in the jet radio structure directly from
the images, we subtracted the image in Fig.~1a from that in
Fig.~1b ($F1b-F1a$) to obtain a differential image (Fig.~7). In
order to minimize the adverse effects of  a limited dynamic range
in the presence of the bright core, we first subtracted the core
component of 1.5~Jy from the two MERLIN data-sets (epochs 1997.14
and 1999.14), re-imaged the residual uv-datasets, and then
produced the differential image presented in Fig.~7.

The largest difference in the flux densities comes from the core.
The peak brightness of the differential image is 0.6 Jy/Beam. Some
variations at a level of 1~mJy per beam in the jet were found.

The increased flux density in the core can be explained as a flare
in the compact unresolved component, as we can see in the light
curve in Fig.~5. Both positive and negative variations are found
above the uncertain set by the noise in the original images (see
Fig.~1a\&b) in the first 0.5$\arcsec$ of the jet (see Fig.~7).
This may be attributed to the continuous movement of the jet. The
variations in the areas of D1 and D3 can not be affected by the
flare from the core, since it would take a few thousand years to
propagate through the distance of 2$\arcsec$ at the speed of
light. This leads to the conclusion that some flares may occur,
independently from the core, in the knot (D3) and hotspot (D1).

\subsection{Spectral index distributions on kpc-scales}

Multi-frequency observations with the VLA and MERLIN  allow us to
produce a spectral index distribution of the source on the kpc
scale (S$_\nu\propto\nu^{\alpha}$). A spectral index distribution
obtained with simultaneous VLA data at 8.5 and 22.5~GHz is
presented in Fig.~8a by aligning the peak emission of the cores.
An upper limit to the opacity shift of 13 mas was estimated for
the optically thin component D3, while a frequency-dependent
position difference of the core of 7 mas was estimated with
Lobanov's model (\cite{lobanov}).

We also present a non-simultaneous spectral index distributions
based on the MERLIN data between 1.6~GHz (epoch 1997.41) and 5~GHz
(epoch 1997.14) in Fig.~8b without consideration of the opacity
shift. An upper limit to the opacity shift of 20 mas was estimated
by comparing the modelfit results of jet component D3, and 7.4 mas
of frequency-dependent position difference of the core was
estimated (\cite{lobanov}).

The spectral index distributions of the source (Fig. 8a\&b) show
that the strongest compact component has a flat spectrum.

A steep spectrum ring appears around the core in Fig.~8a since the
size of the core at 22~GHz is only about half of that at 8.5~GHz.

The kpc scale jets clearly have a steeper spectrum than the core
(see Fig.~8a$\&$b). There is also a general steepening of the
radio emission towards the edges of the jet emission. Some
isolated flat spectrum spots in the jet are also observed,
especially in Fig.~8b. These may be real or due to the
observations being made at different epochs.

Slice plots of the spectral index distributions along the jet axis
at position angle of -18{\degr} are shown in Fig. 9, corresponding
Fig.~8a and 8b, respectively. The slice lines do not intersect the
peak of the each component since the components are not aligned on
a straight line. We can also estimate the spectral indexes with
the modelfit results of Table 3 (see Table ~\ref{si-kpc}).

It is clear that the core has a flat spectral index, while the jet
components D1, D2, , D3 and D4 have steep spectral indexes. Each
component has a steep spectrum edge.

%__________________________________________________ One column table
   \begin{table}
   \centering
      \caption[]{The spectral index on the kpc scale}
         \label{si-kpc}
       \[
     \begin{tabular}{c|ccccc}
     \hline
Bands         &    D    &   D1  & D2    &  D3   & D4      \\\hline
 8.5~-~22.5 GHz  &    0.51 & $-0.7$  &         &  $-0.3$  &         \\
 1.6~-~~5.0 GHz  &    0.04 & $-0.9$  & $-0.78$ &  $-1.1$  & $-0.25$    \\
\hline
\end{tabular}
   \]
\end{table}

\section{Morphology and structural variability on pc-scales}

\subsection{Model fitting of the VLBI data}

\begin{table*}
\centering
 \caption{Model-fitting of the VLBI data} \label{model-pc} %\label{proper}
 \[
\begin{tabular}{c|cccrccr|cccccc} \hline
Obs. inf.   & Comp& $S_{c}$ &   r     & $P.A.$  &  a     &  b/a & $p.a.$ &$\delta_{eq}$&$\delta_{IC}$& $\gamma$ & $\theta$ &$T_{b}$&$T_{r}$\\
            &     &(mJy)    & (mas)   &($\degr$)& (mas)  &     &($\degr$)&             &      &    &($\degr$)&K&K\\
 (1)        & (2) & (3)     &  (4)    & (5)    & (6)     & (7)  & (8)    &   (9)& (10) &(11)  & (12)& (13) &(14)\\ \hline
1996.43     &  C  & 2097    &  0.0    &$  0.0$ & 0.55    & 0.24 &$-30.6$ &  15.8& 15.5 &      &     & 2.5$\times10^{12}$&1.6$\times10^{11}$\\
 VLBA       &  C3 & 141.2   &  3.22   &$ -2.9$ & 1.48    & 1.0  &$  0  $ &      &      & 11.4 & 3.4 & 5.3$\times10^{9} $&\\
 C Band     &  C2 & 133.9   &  4.92   &$ 18.0$ & 1.88    & 1.0  &$  0  $ &      &      & 13.8 & 3.6 & 3.1$\times10^{9} $&\\
            &  C1 & 129.5   &  11.01  &$ 34.9$ & 9.05    & 0.67 &$-14.9$ &      &      &      &     & 2.0$\times10^{8} $&\\
{\it Errors}&     &{\it12\%}&{\it15\%}&{\it 7} &{\it16\%}&      &{\it 7} &      &      &      &     &                   & \\\hline
1997.14     &  C  & 1109    &  0.0    &$  0.0$ & 0.51    & 0.41 &$ -25 $ &  5.0 &  6.0 &      &     & 0.9$\times10^{12}$&1.8$\times10^{11}$\\
 EVN        &  C3 & 117.8   &  3.95   &$ -3.6$ & 1.69    & 1.0  &$ 0   $ &      &      & 13.6 & 8.9 & 3.5$\times10^{9} $& \\
C Band      &  C2 & 172.4   &  5.06   &$ 20.2$ & 3.93    & 1.0  &$ 0   $ &      &      & 21.2 & 7.5 & 9.4$\times10^{8} $&\\
{\it Errors}&     &{\it18\%}&{\it20\%}&{\it 10}&{\it20\%}&      &{\it7}  &      &      &      &     &                   &\\\hline
1997.42     &  C  & 927.3   & 0.0     &$ 0.0 $ & 0.90    & 1.0  &$ 0   $ &  5.3 & 4.3  &      &     & 0.9$\times10^{12}$&1.7$\times10^{11}$\\
 VSOP       &  C3 & 135.0   & 3.01    &$  7.2$ & 2.05    & 1.0  &$ 0   $ &      &      & 13.1 & 8.8 & 2.7$\times10^{10}$&\\
L Band      &  C2 & 240.2   & 5.66    &$ 18.2$ & 4.11    & 1.0  &$ 0   $ &      &      & 20.3 & 7.3 & 1.2$\times10^{10} $&\\
            &  C1 & 104.4   &10.97    &$ 35.5$ & 5.73    & 1.0  &$ 0   $ &      &      &      &     & 2.7$\times10^{9} $&\\
{\it Errors}&     &{\it17\%}&{\it15\%}&{\it 10}&{\it20\%}&      &{\it7}  &      &      &      &     &                   & \\\hline
1998.12     &  C  & 504.0   & 0.0     &$ 0.0 $ & 1.0     & 0.39 &$ 1.4 $ &  0.5 & 0.9  &      &     & 1.1$\times10^{12}$& ... \\
 Global     &  C3 &  89.9   & 3.71    &$ -1.6$ & 1.75    & 1.0  &$  0  $ &      &      &      &     & 2.5$\times10^{9} $&\\
 C Band     &  C2 &  95.4   & 5.59    &$ 21.1$ & 2.66    & 1.0  &$  0  $ &      &      &      &     & 1.1$\times10^{9} $&\\
            &  C1 &  82.6   &10.10    &$ 33.1$ & 7.84    & 1.0  &$  0  $ &      &      &      &     & 1.1$\times10^{8} $&\\
{\it Errors}&     &{\it18\%}&{\it10\%}&{\it 10}&{\it10\%}&      &{\it7}  &      &      &      &     &                   &\\\hline
1999.01     &  C  & 2245    & 0.0     &$ 0.0 $ & 0.15    & 0.37 &$15.4 $ & 17.1 & 22.4 &      &     & 2.5$\times10^{12}$&1.4$\times10^{11}$ \\
 VLBA       &  C4 &  26.3   & 1.25    &$ -4.6$ & 0.8     & 1.0  &$ 0   $ &      &      & 12.7 & 3.3 & 3.8$\times10^{8} $&\\
 U Band     &  C3 &  17.7   & 4.19    &$ -2.6$ & 1.5     & 1.0  &$ 0   $ &      &      & 11.8 & 3.0 & 7.4$\times10^{7} $&\\
            &  C2 &  28.0   & 6.23    &$ 16.0$ & 2.6     & 1.0  &$ 0   $ &      &      & 14.0 & 3.1 & 3.8$\times10^{7} $&\\
{\it Errors}&     &{\it15\%}&{\it13\%}&{\it 5} &{\it13\%}&      &{\it7}  &      &      &      &     &                   &\\\hline
1999.14     &  C  & 1387    & 0.0     &$  0.0$ & 0.39    & 0.40 &$ 3.4 $ & 12.3 & 11.8 &      &     & 2.0$\times10^{12}$&1.6$\times10^{11}$\\
 EVN        &  C4 & 42.2    & 1.55    &$ -9.6$ & 1.1     & 1.0  &$ 0   $ &      &      & 11.9 & 4.6 & 2.9$\times10^{9} $&\\
 C Band     &  C3 & 31.9    & 4.26    &$ -5.7$ & 1.2     & 1.0  &$ 0   $ &      &      & 10.6 & 4.6 & 1.8$\times10^{9} $&\\
            &  C2 & 76.3    & 5.83    &$ 15.8$ & 2.3     & 1.0  &$ 0   $ &      &      & 13.7 & 4.6 & 1.2$\times10^{9} $& \\
            &  C1 & 71.7    &10.62    &$ 30.0$ & 6.6     & 1.0  &$ 0   $ &      &      &      &     & 1.3$\times10^{8} $& \\
{\it Errors}&     &{\it15\%}&{\it15\%}&{\it 6} &{\it15\%}&      &{\it7}  &      &      &      &     &                   &\\\hline
1999.45     &  C  & 1765    &  0.0    &$ 0.0 $ & 0.35    & 0.44 &$ -6.9$ & 18.5 & 12.4 &      &     & 2.7$\times10^{12}$&1.4$\times10^{11}$\\
 Global     &  C4 &  52.9   &  1.55   &$ -6.5$ & 1.33    & 1.0  &$   0 $ &      &      & 13.1 & 2.8 & 2.4$\times10^{9} $&\\
C Band      &  C3 &  54.1   &  4.31   &$ -3.7$ & 2.06    & 1.0  &$   0 $ &      &      & 12.3 & 2.7 & 1.1$\times10^{9} $&\\
            &  C2 & 108.5   &  6.14   &$ 20.0$ & 3.55    & 1.0  &$   0 $ &      &      & 14.3 & 3.0 & 7.4$\times10^{8} $&\\
            &  C1 &  92.1   & 11.2    &$ 33.3$ & 8.07    & 1.0  &$   0 $ &      &      &      &     & 1.2$\times10^{8} $&  \\
{\it Errors}&     &{\it 9\%}&{\it10\%}&{\it 5} &{\it10\%}&      &{\it 5} &      &      &      &     &                   &\\\hline
2000.12     &  C  & 931.3   & 0.0     &$ 0.0 $ & 1.6     & 0.28 &$-19.1$ &  6.1 & 4.8  &      &     & 1.1$\times10^{12}$& 1.8$\times10^{11}$ \\
  VLBA      &C2\&3& 245.1   &5.23     &$ 16.4$ & 3.9     & 1.0  &$ 0   $ &      &      & 15.1 & 7.5 & 1.3$\times10^{10} $&  \\
L Band      &  C1 & 195.0   &10.53    &$ 31.2$ & 7.9     & 1.0  &$ 0   $ &      &      &      &     & 2.5$\times10^{9 }$& \\
            &  C0 &  88.1   &31.66    &$ -1.1$ &25.4     & 1.0  &$ 0   $ &      &      &      &     &                   & \\
{\it Errors}&     &{\it10\%}&{\it12\%}&{\it 5} &{\it13\%}&      &{\it 5} &      &      &      &     &                   &\\\hline
\end{tabular}
\]
\end{table*}

The VLBI self-calibrated data-sets associated with each image in
Fig.~3 were fitted with Gaussian components using DIFMAP. An
elliptical Gaussian component was fitted to the strongest
component, and circular Gaussian brightness distributions were
used for the jet components to estimate their sizes.
%in particular, to measure the opening angle of the jet.

The parameters of model fitting are listed in
Table~\ref{model-pc}. Column 1 gives the observation epoch, array,
and frequency. columns 2 to 8 are the same as the columns 3 to 9
in Table~\ref{model-kpc}. The uncertainties of the model fitting
are estimated in the same way as for the acrsecond-scale images
(see section 4.1) and are listed in the table.

Five components are fitted to the VLBI data.  In order to compare
our results with previous publications, we label the components
following the convention introduced by Piner and Kingham (1997)
from C to C4 (see Fig.~10).

Fig.~10 is the model-fitted image (Gaussian components restored
with a $2.4\times1.1$~mas, at $p.a. -2.1\degr$ beam) of 1156+295
based on the Global VLBI data at the epoch 1999.45 at 5~GHz. This
is in good agreement with the CLEANed image shown in Fig.~3g.

The core contributes a higher fraction of the total flux density
at higher frequencies. The fraction of the core flux density is
about 97$\%$ at 15~GHz, 80 - 86$\%$ at 5~GHz (except the lowest
level of 65$\%$ at the minimum of activity at the epoch 1998.12),
and 65 - 70$\%$ at 1.6~GHz. This is a clear indication that the
high frequency emission comes from the inner regions of the
source. The fraction of the core flux density is higher during
active outburst phases (e.g. 85$\%$ at the epoch 1999.45 at 5~GHz)
compared to more quiescent epochs (e.g. 65$\%$ at the epoch
1998.12 at 5 ~GHz). This suggests that most of the flare occurred
in an area less than 1 mas in size.

For the individual components, we note: a) we use one elliptic
Gaussian component C1 to fit the two separated outer emission
regions in the north-east area at epoch 1996.43 (Fig.~3a); b) C1
is not detected at the epoch 1997.14 due to sensitivity
limitations and is resolved at 15 GHz (epoch 1999.01); c)
components C2 and C3 are not resolved in the 1.6 GHz VLBA
observations and are fitted as a combined centroid component
C2\&3; d) C4 is detected after 1999 at 15~GHz; e) outer component
C0 is detected at a distance of 30 mas from the core (Fig.~3h).

\begin{figure}
\vspace{150pt} \includegraphics{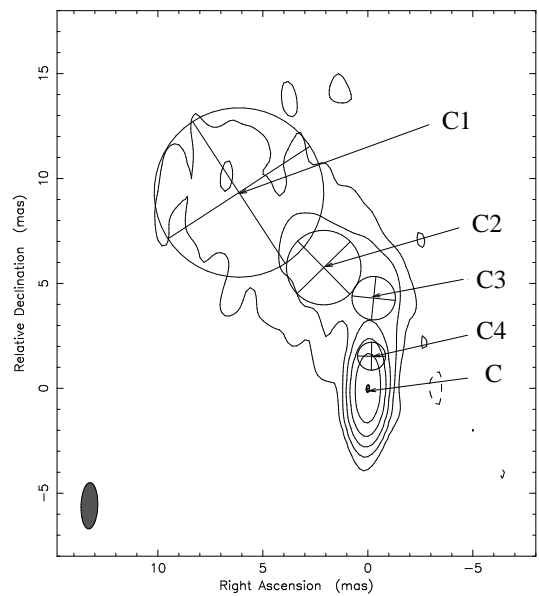}
  \vspace{80pt} \caption{The model image at the epoch 1999.45 at 5~GHz.
Peak brightness is 1.71~Jy/beam. Contour levels are 1.5~mJy/beam
$\times (-1,1, 1, 4, 16, 64, 256)$.}
\end{figure}

\subsection{The differential VLBI image at 5~GHz}

As noted in section 5.1, the largest structural variations during
the flare appeared in the vicinity of the core. The two high
sensitivity global VLBI observations at 5~GHz (epochs 1998.12 and
1999.45) were made in the states of low and high flaring activity
respectively.  The differential image of these two global VLBI
datasets ($F3g - F3d$) is shown in Fig.~11. Most of the different
flux density in the image comes from the core area. The peak
(residual) flux density is 1.3 Jy/beam, which is about 3 times of
that of the total intensity peak observed at the epoch 1998.12.
This large difference in the core region is clearly associated
with the violent flare in the core that has occurred after the
epoch 1998.12 and during the epoch 1999.45.

The existence of a `negative emission' component in the
differential image (Fig. 11) can be explained by a variation in
the direction of the jet. When the direction of the relativistic
jet comes close to the line of sight, its flux density is enhanced
by a Doppler boosting; when the orientation of jet changes from
the line of sight, the component becomes fainter. This pattern is
characteristic of a relativistic jet with a helical trajectory.

{The observed feature can be also explained directly by the proper
motion of the jet. If one component moves from point A to point B,
an observer will see a negative component at the point A and a
positive component at point B with the same absolute value of flux
density. This seems not like in the case of Fig.~11.}

\begin{figure}
\vspace{150pt}
 \includegraphics{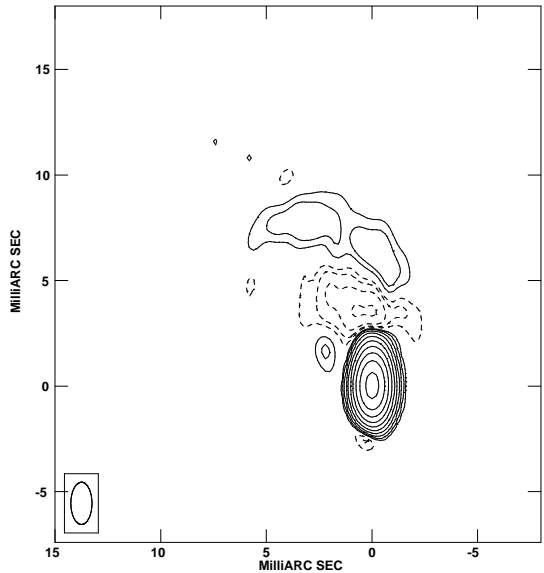} \vspace{80pt} \caption{The differential
image of the two 5 GHz Global VLBI data-sets from the epochs
1999.45 and 1998.12, the peak brightness is 1.3~Jy/beam., contour
levels are 2.0~mJy/beam $\times (-8, -4, -2, -1, 1, 2, 4, 8, 16,
32, ..., 512)$. FWHM: 2 $\times$ 1 mas, at $0\degr$}
\end{figure}

\subsection{Proper motion of the jet components}

\begin{figure}
\vspace{85pt} \includegraphics{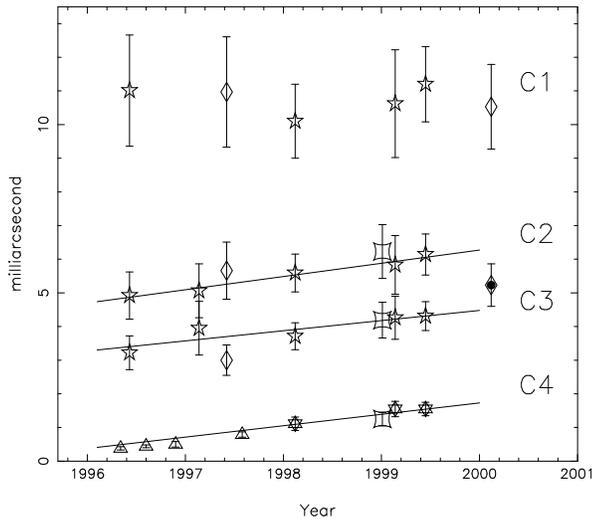} \vspace{130pt} \caption{Apparent proper
motion of the jet components: 1.6~GHz (diamond), 5~GHz (star),
15~GHz (square), 22~GHz (triangle, Jorstad, et al. 2001). }
\end{figure}

The positions of the components during the period from 1996 to
2000 are shown in Fig.~12. The x-axis is the time and the y-axis
shows the distance of the components from the core. Different
symbols are used to represent different observing frequencies: a
diamond for 1.6~GHz, a star for 5~GHz, and a square for 15~GHz. We
also plot a triangle for the 22~GHz component labelled as B2 by
Jorstad et al. (2001). We identify B2 is the same with the
component C4 in this paper (see Table 4, Fig.~10). The components
C2 and C3 are not resolved at 1.6~GHz at the epoch 2000.12. This
combined centroid is shown as a filled diamond symbol.

Since the positions of the components could be frequency
dependent, we estimate the proper motion of components C2, C3 and
C4 via the 5 GHz data sets. These are 0.39$\pm$0.1, $0.30\pm0.08$
and $0.34\pm0.08$ mas/yr respectively. These values correspond to
the apparent velocities of $13.7\pm3.5, 10.6\pm2.8$, and
$11.8\pm2.8c$.

The component C1 is located at the point where the jet turns
sharply. It does not appear to have any appreciable motion in the
r direction, while the position angle changed from about 35\degr
to 31\degr, which correspond to a apparent velocity of about 7$c$.

Jorstad et al. (2001) reported the proper motion measured at 22
GHz of three components in this source, the period of the VLBA
observation is overlapped with ours, the reported apparent
velocity of B2 (labelled as C4 in this paper) is $11.8\pm1.2$
(corrected to $q_0=0.5$), based on three epochs during period
1996.60 to 1997.58, which is in agreement with our result. The B3
component in their paper is unresolved in our VLBI data due to the
limited resolution at low frequencies. Piner \& Kingham (1997)
reported that the apparent velocities of C1 to C4 are $13.5\pm3.5,
8.1\pm1.7, 8.5\pm1.4$, and $5.4\pm1.8~c$, respectively, based on
their geodetic data obtained during the period 1988.98 to 1996.25.

The apparent radio component velocity of 40$c$ is the highest
super-luminal velocity reported for AGN to date (\cite{M1};
\cite{M2}). However, as this value was derived based on four-epoch
of VLBI observations at three frequencies, the apparent motion
could be 'caused' by frequency-dependent effects.

\subsection{Spectral index distribution on pc-scales}

\begin{figure*}
\vspace{150pt} \includegraphics{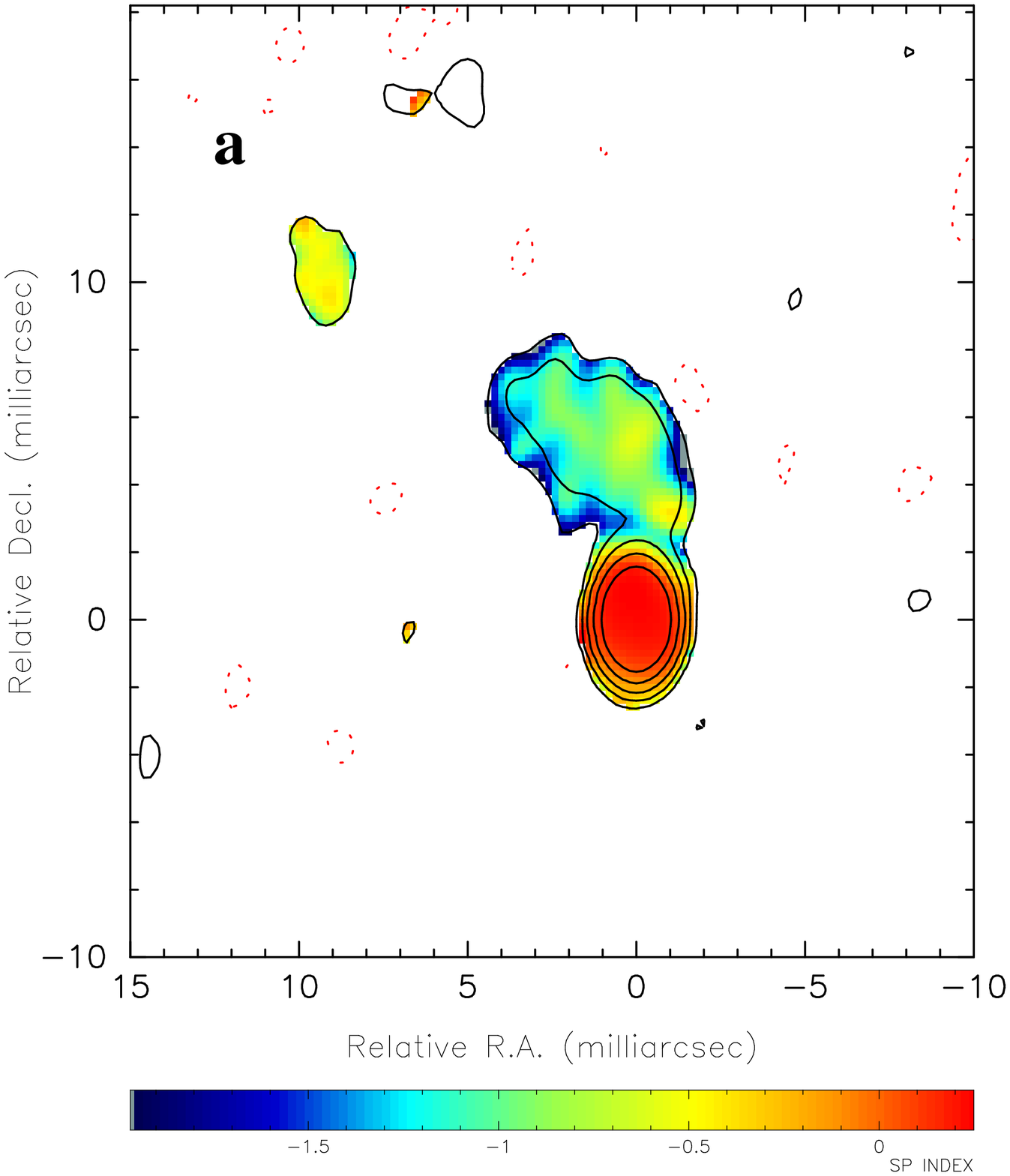} \includegraphics{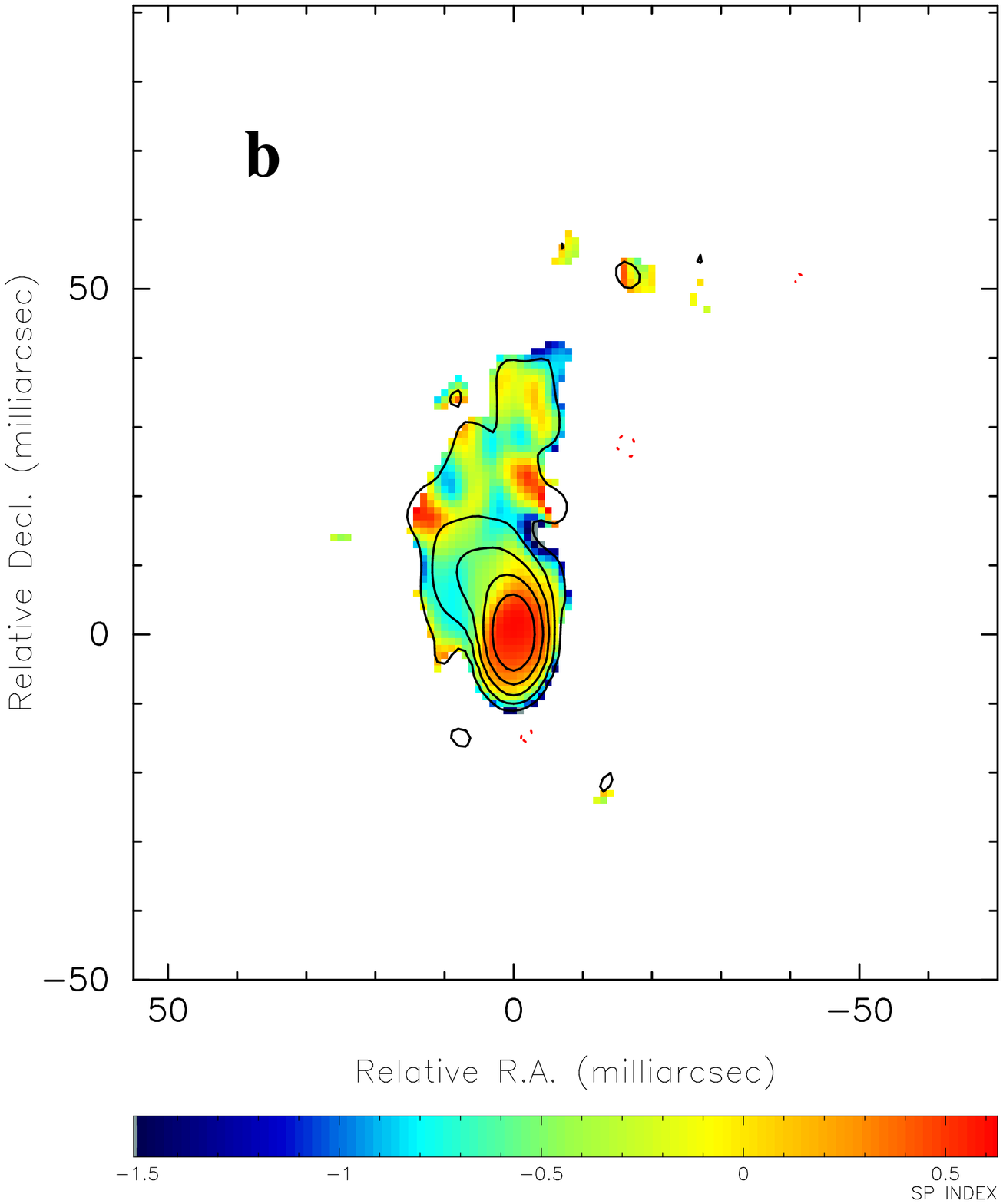} \vspace{80pt} \caption{(a):
The 5 $vs$ 15~GHz spectral index distribution (color),
superimposed on the 15~GHz VLBA image at the epoch 1999.01, the
wedge at the bottom shows the spectral index range $-1.95$ to
0.25; (b): the 1.6 $vs$ 5~GHz spectral index distribution (color),
superimposed on the 5~GHz VLBA image, the wedge on the bottom
shows the spectral index range $-1.5$ to 0.62. The contours levels
in a~\&~b are 2 mJy/beam $\times (-1, 1, 4, 16, 64, 256)$ with
FWHM of 1.5$\times$1 mas and 7$\times$4 mas, respectively.}
\end{figure*}

Although no simultaneous VLBI data of the source at different
frequencies are yet available, we attempted to estimate the
spectral index distribution of the jet on pc scales using the
multi-epoch/multi-frequency images presented here. The
non-simultaneous spectral index distribution between 1.6 (epoch
2000.15) and 5~GHz (epoch 1999.45), as well as between 5 (epoch
1999.45) and 15~GHz (1999.01) are presented in Fig.~13 by summing
the zero opacity shift in the cores. The upper limit opacity shift
of 0.3 and 0.7 mas were estimated by comparing the modelfit
results of the optical thin jet component for Figs.~13a and 13b,
respectively. Meanwhile, 0.5 mas of the uncertainty of the
frequency-dependent position difference of the core was estimated
with Lobanov's model (\cite{lobanov}) for the spectral index maps
Figs.~13a and 13b.

Fig.~13a shows the 5$-$15~GHz spectral index distribution. It is
clear that the core area has a flat spectrum $\alpha \sim 0.25$.
The emission becomes a steep spectrum ($\alpha \sim -0.7$ to
$-1.5$) outwards from the core.

Fig.~13b is the 1.6$-$5~GHz spectral index distribution. There is
an inverse spectrum ($\alpha \sim 0.58$) in the core area.  Two
flat spectrum features are seen in the jet.

Figs.~14a and 14b show the profile of the 5$-$15~GHz spectral
index along the direction of P.A.=0{\degr} and $1.6-5$~GHz
spectral index along P.A.= 25{\degr}, corresponding to Figs. 13a
and 13b, respectively.

Table~\ref{si-pc} gives the spectral index values estimated from
the model-fit components listed in  Table~\ref{si-pc}.

%__________________________________________________ One column table
   \begin{table}
   \centering
      \caption[]{The spectral index on the pc scale}
         \label{si-pc}
       \[
     \begin{tabular}{c|ccccc}
     \hline
Bands         &    C    &   C1  & C2    &  C3   & C4 \\\hline
 5~-~15 GHz   &    0.22 &       & $-1.2$& $-1.0$& $-0.6$        \\
 1.6~-~5~GHz  &    0.58 & $-0.7$  &   &     &      \\\hline
\end{tabular}
   \]
\end{table}

\begin{figure*}
\vspace{195pt}
 \includegraphics{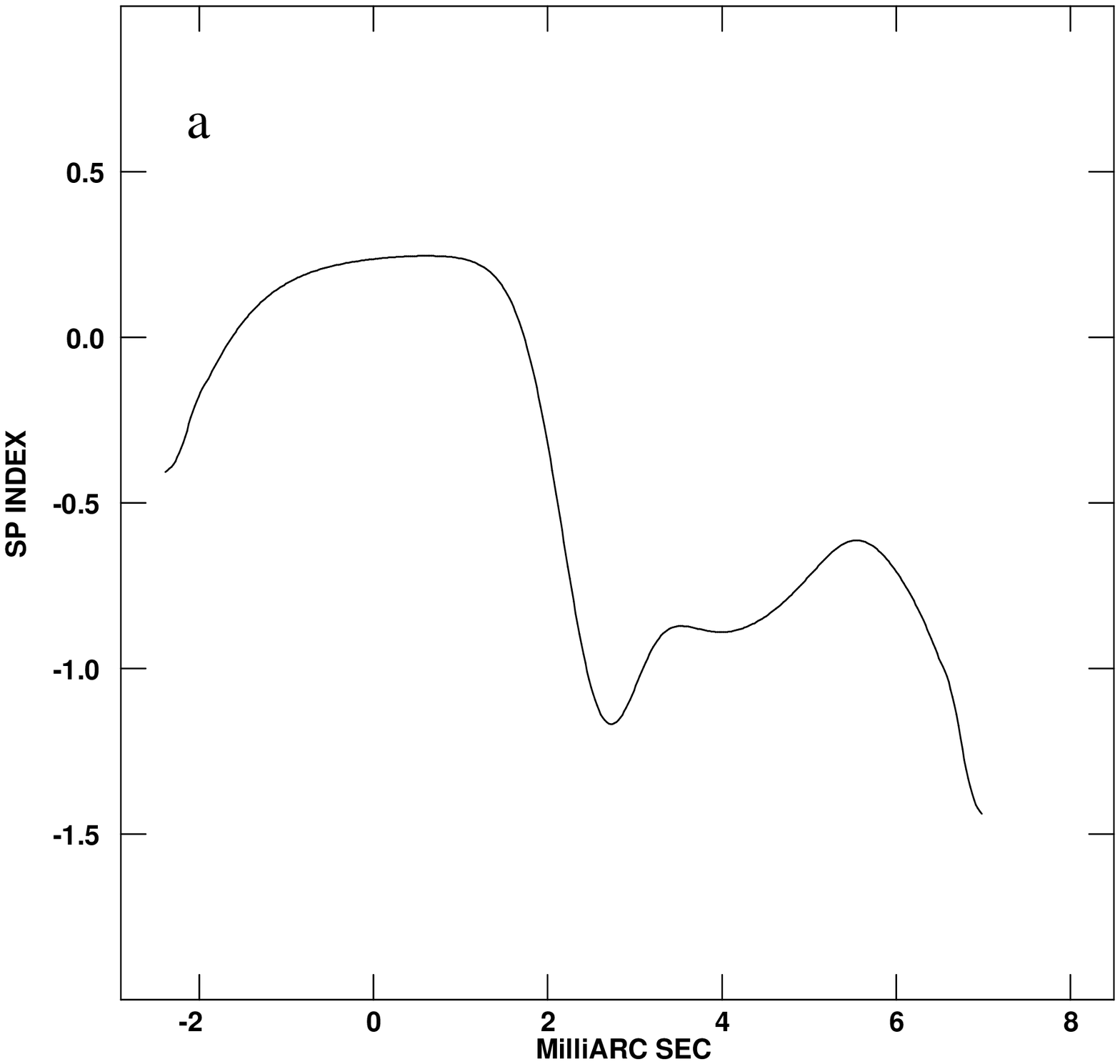} \includegraphics{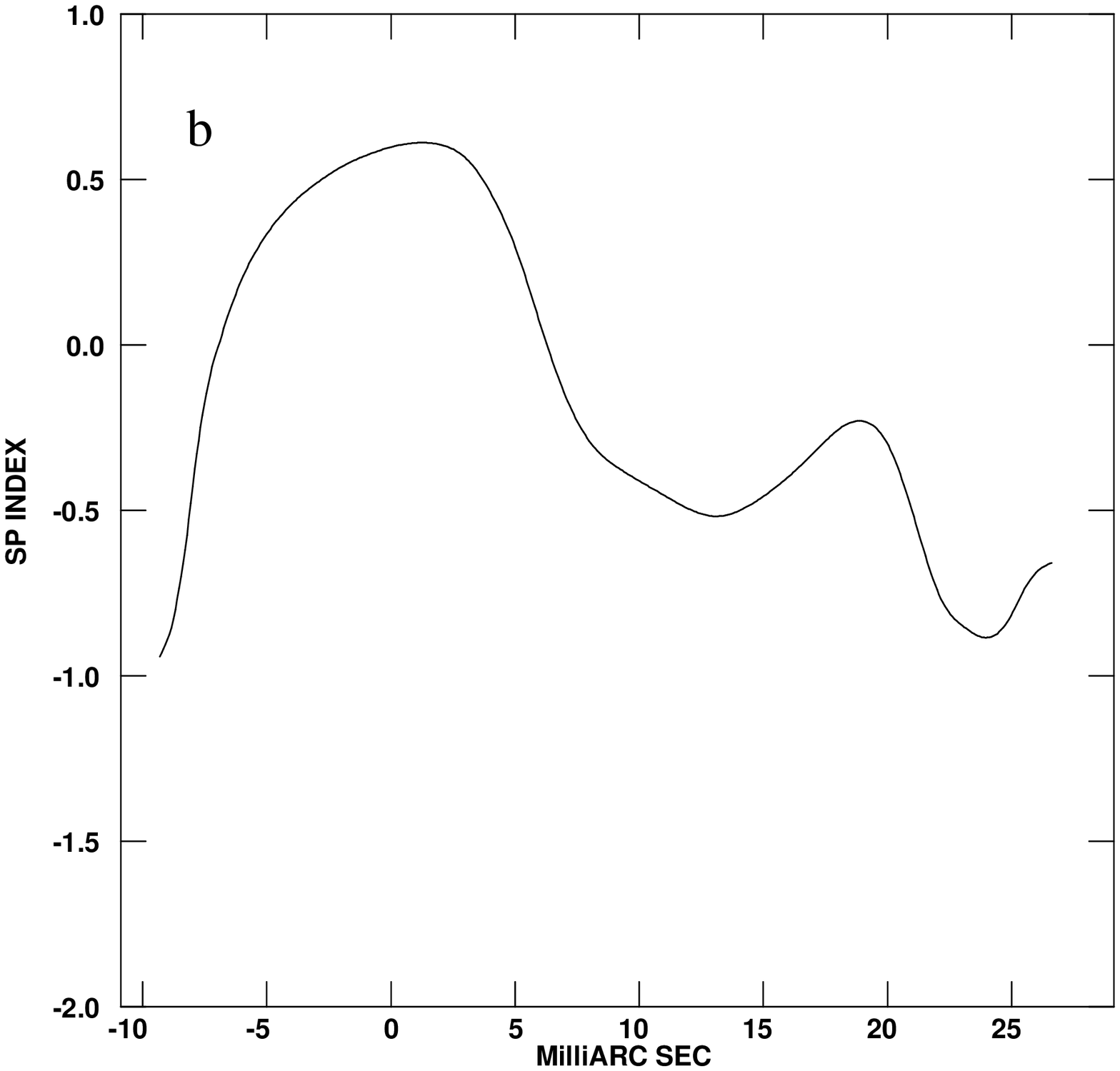} \vspace{50pt} \caption{(a):
Slice plot of the spectral index at p.a. 0{\degr} corresponding to
Fig. 13a; (b): Slice plot of the spectral index at p.a. 25{\degr}
corresponding to Fig.13b}
\end{figure*}

\section{Physical parameters of the core and jet}

\subsection{The Equipartition Doppler factor $\delta_{eq}$}

As we mentioned in the previous sections, the flares of 1156+295
may be enhanced by the Doppler boosting. The Doppler factor of the
outflow from compact radio cores can be estimated assuming the
energy equipartition between the radiating particles and the
magnetic field (\cite{Read94}; \cite{G2}).

The method was considered using the radio observed parameters at
the turnover frequency, but actual VLBI observations were not done
at the turnover frequency. We will use the radio parameters we
obtained from our VLBI observations to roughly estimate the
Doppler boosting in 1156+295.

By following the assumptions of Scott $\&$ Readhead (1977) and
Marscher (1987), the Equipartition Doppler factor $\delta_{eq}$ of
the core component for each epoch is estimated and listed in
column 9 of Table~\ref{model-pc}. The values are in the range of 5
to 18.5 except 0.5 at the epoch 1998.12. The latter is expected,
since the source is at its lowest state.

\subsection{Inverse Compton Doppler factors $\delta_{IC}$}

For comparison, the Doppler factor of the core component can also
be estimated with the Synchrotron self-Compton emission in a
uniform spherical model (\cite{Mar87}).

Under the assumption in section 6.1, we estimated $\delta_{IC}$
using the equation (1) of Ghisellini et al. (1993) with the
$0.075~\mu$Jy X-ray flux density of 1156+295 at 2~keV (\cite{M0}).
The inverse Compton Doppler factors $\delta_{IC}$  of 1156+295 at
all epochs were listed in the column 10 of Table~\ref{model-kpc}.
The values cover a wider range from 4.3 to 22.4 except 0.9 at the
epoch 1998.12.

The estimated values may be a lower limit since part of X-ray
emission could come from the thermal or other emission. The
variability of X-ray flux and  non-simultaneous observations
between radio and X-ray are causes of uncertainty in estimating
$\delta_{IC}$.

The $\delta_{IC}$ and the $\delta_{eq}$ are comparable. The
average values of $\delta_{IC}$ is about 96\% of that of
$\delta_{eq}$. The Doppler factor varied with time. The variation
was correlated with the total flux density variation of the
source. The core has higher Doppler factor at a higher frequency
than that at lower frequencies.

\subsection{Lorentz factor and viewing angle}

In the relativistic beaming model $\beta_{app} $ depends on the
true $\beta(v/c)$ factor and the angle to the line of sight
$\theta$ (\cite{R0}),
\begin{equation}
\beta_{app}=\frac{\beta\sin\theta}{1-\beta\cos\theta}\;\;.
\end{equation}
The Doppler factor can be written as
\begin{equation}
\delta = \gamma^{-1}(1-\beta\cos\theta)^{-1}\;\;,
\end{equation}
\noindent where $\gamma =1/\sqrt{1-\beta^{2}}$.

The Lorentz factor $\gamma$ and the viewing angle $\varphi$ can be
computed (Ghisellini et al. 1993):
\begin{equation}
\gamma=\frac{\beta_{app}^2+\delta^2+1}{2\delta}\;\;,
\end{equation}
\begin{equation}
\tan \varphi=\frac{2\beta_{app}}{\beta_{obs}^2+\delta^2-1}\;\;.
\end{equation}

The $\gamma$ and $\phi$ of C2, C3, and C4 shown in columns 11 and
12 of Table 4 are estimated from the Doppler factor
($\delta=\delta_{eq}$ is used) and their apparent velocities
$\beta_{app}$. The Lorentz factor of each epoch changes between
13.8 and 21.2, 10.6 and 13.6, 11.9 and 13.1 for C2, C3 and C4,
respectively, while the viewing angles varied in the range of
2.7$\degr$ to 8.9$\degr$.

Ghisellini et al. reported that the jet to counter-jet brightness
ratio can be estimated from (1993):

\begin{equation}
J=(\frac{1+\beta\cos\theta}{1-\beta\cos\theta})^{2-\alpha}\;\;.
\end{equation}

Taking $J_{kpc-jet}=5.36$ and $J_{lobe}=1.66.$ (see section 4.1),
and assuming the spectral index of the kpc jet and the lobe
$\alpha=-0.9$ (Table~\ref{si-kpc}), we have:

\begin{equation}
(\beta\cos\theta)_{kpc-jet}=0.3\;\;,
\end{equation}

\begin{equation}
(\beta\cos\theta)_{lobe}=0.1\;\;.
\end{equation}

It indicates that the kpc jet is faster and/or has a smaller
viewing angle than the lobe components.

\subsection{The brightness temperature}

A high brightness temperature in flat spectrum radio sources is
usually considered to be an argument supporting relativistic
beaming effect. In the synchrotron model, with the brightness
temperature close to $\sim 10^{12}$~K, the radiation energy
density dominates the magnetic energy density, and produces a
large amount of energy in a very short time scale. The
relativistic particles quickly cool (Compton catastrophe). The
maximum brightness temperature that can be maintained is about
$\sim 10^{12}$~K in the rest frame of the emitting plasma
(\cite{KP}).

\begin{figure}
\vspace{80pt} \includegraphics{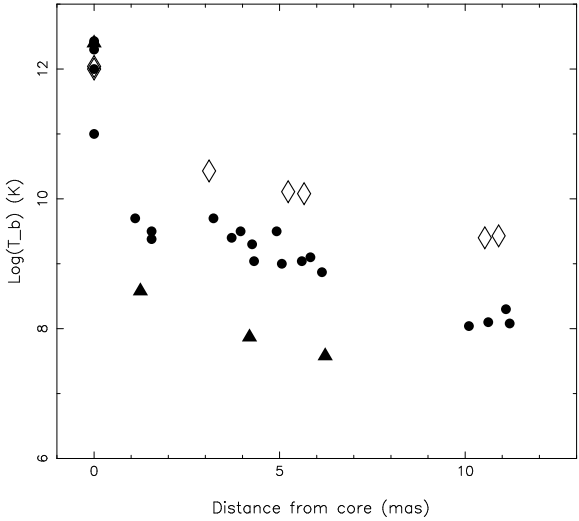} \vspace{125pt} \caption{The brightness
temperature distribution of VLBI components: 1.6~GHz diamond,
5~GHz circle, 15~GHz triangle.}
\end{figure}

The brightness temperature $T_b$ of an elliptical Gaussian
component in the rest frame of the source is (\cite{Shen97})

\begin{equation}
T_b = 1.22 \times 10^{12} \frac{S_{\nu}}{\nu^{2}_{ob}ab}(1+z)
~$K$\;\;,
\end{equation}
where $z$ is the redshift, $S_{\nu}$ is the observed peak flux
density in Jy at the observed frequency $\nu_{ob}$ in GHz, $a$ and
$b$ are the major and minor axes in mas.

The brightness temperatures of the the components of 1156+295 are
calculated with equation (8) at various epochs. The results are
listed in column 13 of Table 4. We present the brightness
temperature distribution in Fig.~15. It is clear that the
component brightness temperatures are inversely proportional to
their observed frequencies. The brightness temperature of the core
component is around $10^{12}$~K, while the brightness temperatures
of the jet components are much lower (around $10^{7-10}$~K).

By considering the Doppler boosting of the source, the intrinsic
brightness temperature $T_r$ can be estimated as $T_r =
T_b/\delta$ and listed in column 14 of Table 4 ($\delta_{eq}$ is
used). The intrinsic brightness temperatures of the core component
at all epochs are almost constant ($\sim~10^{11}$K). This may
serve as an upper limit of the brightness temperature in 1156+295
in the rest-frame of the source. Since the estimated $\delta_{eq}$
is an upper limit, the intrinsic brightness temperatures here are
the lower limit.

\section{A helical pattern}

The helical jet model for AGNs was proposed (e.g. \cite{C1}) to
explain a bi-modal distribution of the difference in the
orientation of arcsecond and mas structures in core-dominated
radio sources. Helical jets are a natural result of precession of
the base of the jet (e.g. Linfield 1981) or fluid-dynamical
instabilities in the interface between the jet material and the
surrounding medium (Hardee 1987). Assuming the initial angular
momentum of the jet is due to the precession of the accretion
disk, the helical structure observed might be related to  the
precession of a wobbling disk while the interface instability
produces a wiggling structure that reflects a wave pattern (e.g.
\cite{zhao}).

As we mentioned in section 3.3, the trajectory of the jet in
1156+295 could be a helix on the surface of a cone (Fig.~6). Fig.
16 shows the measured position angles of all components at all
frequencies and all epochs for 1156+295. The two components (B2
and D) from the Jorstad et al's paper (2001) are also included.
The x-axis shows the distance of the components from the core in
logarithmic scale and the y-axis shows the position angle. This
symmetric distribution of the position angle suggests that the jet
in 1156+295 is spatially curved and the jet components follow a
helical path.

\begin{figure} \vspace{80pt}
%\vspace{250pt}
\includegraphics{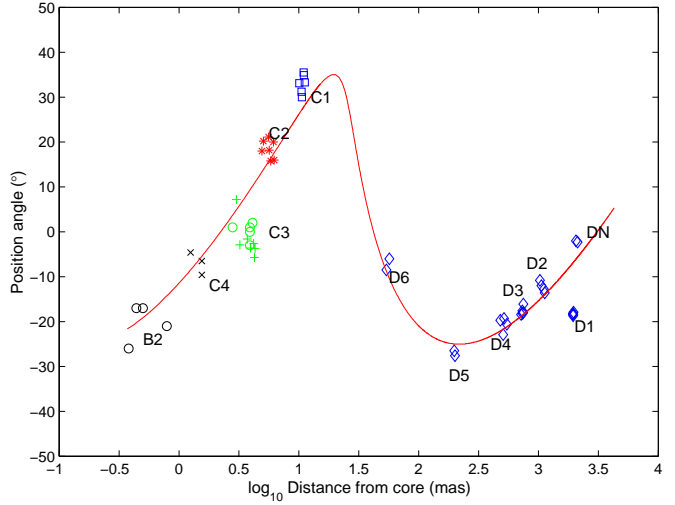} \vspace{125pt} \caption{Measured positions of all
components at all frequencies and all epochs for 1156+295.}
\end{figure}

By assuming the conservation laws for kinetic energy and momentum,
the helical model parameters can be derived following
(\cite{stef95}):

\begin{equation}
z(t) = \beta_{z} t + z_{0}\;\;,
\end{equation}

\begin{equation}
r(t) = \beta_{z} t \;\;{\rm tan}\psi + r_{0}\;\;,
\end{equation}

\begin{equation}
\phi(t) = \phi_{0} + \frac{\omega_{0}r_{0}}{\beta_{z}\;\;{\rm
tan}\psi} \;\;{\rm ln}\frac{r(t)}{r_{0}}\;\;,
\end{equation}
where $z, r, \phi$ are the cylindrical coordinates; $\beta$ is the
velocity of the jet component in units
 of the speed of light;
$\psi$ is opening half-angle of the cone; $t$ is the time in the
rest frame of the source (jet); and $\omega$ is the angular
rotation velocity ($d\phi/d t$).

The estimated viewing angles $\theta$ of inner jet components in
1156+295 are 3$\degr$ to 8$\degr$ (see table 4). A value of
$5\degr$ was adopted as the angle between the line of sight and
the axis of the cone; the half-opening angle of the cone $\psi$=
2.5$\degr$ was estimated from the observed oscillatory amplitude
of position angle. Figs. 16 and 17 show one possible approximation
rather than the best fit to the data.  Fig. 16 represents the
relationship between the position angle and the projected distance
from the core of this helical jet. Fig. 17 shows the projected
path of this helical trajectory in the image plane on the two
different scales. These figures show that a simple helical
trajectory can well represent the radio structure from 1 mas to 1
arcsecond (D2 component).

Beyond the D2 component, the helical path is probably interrupted
or it changes direction due to the interaction between the jet and
the medium. The model does not consider scales less than 1 mas
because of the lack of observational data. Jorstad et al. (2001)
pointed out that the trajectories of components can fall anywhere
in the region of position angle from $-25\degr$ to 25$\degr$ out
to a distance of $\sim$ 1 mas from the core based on their data at
22~GHz. This does not contradict the helical model presented here.

Fig.~18 shows that the components move along curved path. For
comparison, we also included the data from the geodetic VLBI
observation by Piner \& Kingham (1998) (estimates from their Fig.
10) during period from 1988 to 1996 in Fig.~18, labelled as PKC1,
PKC2 and PKC3. The two data sets show the movement of jet
components along a curved trajectory on the time scale of 20
years.

\begin{figure}
\vspace{80pt} \includegraphics{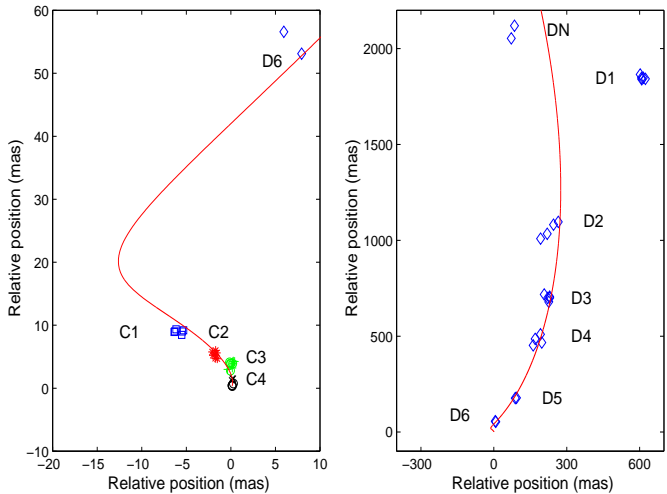} \vspace{120pt} \caption{The positions of
jet components and the projective path of a helical jet in the
image plane on the pc-({\it left}) and the kpc-({\it right})
scales. }
\end{figure}

\begin{figure}
\vspace{80pt} \includegraphics{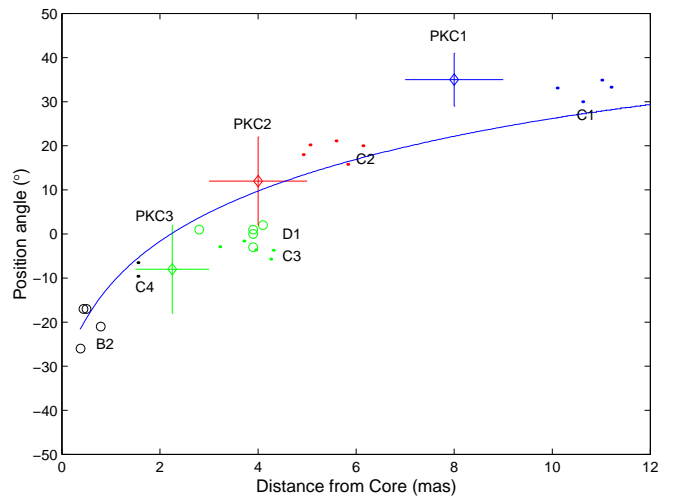} \vspace{120pt} \caption{The evidence for
the movement of the components along a curved path.}
\end{figure}

In addition, we note some small structural oscillations observed
on the mas scale (e.g. components C3 and C1). It is not clear
whether this is a processing on a short time scale or measurement
errors.

If the jet moves on the surface of a cone along a helical
trajectory, the helical wavelength ($\lambda$, the pitch distance)
of this helical trajectory increases with the time. The apparent
period of this helical jet (starting at about 1 mas from the core)
at the projected distance is about 1200 mas. If we assume that the
averaged proper motion ($\mu$ = 0.3 mas/yr) is constant on the 1
mas to 1 arcsecond scale (as we assumed), the lower limit of the
helical period in time in the source-rest frame is about $10^{4}$
years ($P = (1+z)*\lambda$/$\mu$). Considering the possible
deceleration of the jet on the large scale, the real period is
likely to be longer.

The helical jet may be due to a binary black hole in the nucleus
of the source, such as in the  BL Lac Object OJ~287 (e.g.
\cite{Sill}, \cite{LV}). The model for both the optical and radio
light curves observed in OJ~287 suggests that the helical jet and
double peaked cyclic outbursts can be produced by a pair of
super-massive black holes in a binary system (\cite{villa}) if the
twin jets ejected from each of the BHs were in a precession around
the orbital axis of the binary and the Doppler boosting effect is
important.

Alternatively, in a spinning super-massive black hole system, if
the disk axis offsets from the spin axis, the intrinsic torque of
the spin black hole placing on an accretion disk could also
produce a precession of the axis of the accretion  disk
(\cite{LM}). The variation of the direction in the jet motion
produced from a wobbling disk can also be responsible for the
helical pattern.

\section{Summary}
We investigated the radio structures of the radio-loud quasar
1156+295 in detail on various angular scales using different
interferometric arrays at different frequencies.

1. The source shows an asymmetric double structure at
 l.6~GHz with 250 mas resolution (about 1.5 kpc),  the northern jet is
 relativistically Doppler boosted, the southern emission is resolved into
 several low brightness regions (see the color figure of the cover).

2. At 80 mas resolution (about 0.5 kpc), the northern jet has an
 almost straight structure at $P.A.=-18\degr$ with some sinusoidal
 fluctuations. The source is dominated by a flat core, and the jet
 emission {has a steep spectrum}.

3. On the VLBI scale, the jet bends from the north to the
north-east at
 3$\sim$4 mas from the core, and then turns about $90\degr$ to
 the north-west, thus aligning with the direction of the arcsecond-scale jet.
 The 1.6~GHz VLBA image links the VLBI emission and arcsecond
 emission of the jet, which follows an S type structure. This may
 be an evidence of a helical jet.
 The spectral index distribution at the mas scale shows a flat spectrum VLBI
 core and a steep spectrum jet. The differential image demonstrates that
 the flux density variation mainly comes from the very compact core area.

4. The proper motions in three components were detected; the
 apparent superluminal velocities are in the range of $ 10.6$
 to $13.7~c$. The differential images also show evidence of bulk motion
 of the emitting material along a  curved trajectory.

5. Ejection of the VLBI components corresponds to the flares in
the light curves. This is consistent with the hypothesis that
total flux density variations mainly occur in the core region.

6. High Doppler factors ($\sim$18), and a high Lorentz factor
($\sim$21) were measured in the northern jet components on
pc-scales supporting the assumption that the jet moves
relativistically.

7. We found that a helical trajectory along the
 surface of a cone can well represent the radio structure from 1 mas
 to 1000 mas, based on the conservation laws for kinetic energy
 and momentum. The period of the helical
 path is longer than $10^{4}$ years.

The study provides the most comprehensive account so far of kpc-
and pc-scale morphological properties of the radio emission in the
AGN 1156+295. These properties are analyzed in the framework of
the jet helical model  e.g. a binary black hole in the nucleus of
the source. However, other models, e.g. based on the
Kelvin-Helmholz instabilities (\cite{H1}), the trailing shock
model in a relativistic jet (\cite{gomez}) or MHD processes could
be verified in future studies against the observing data
presented.

%Further multi-epoch VLBI observations at higher frequencies (2cm,
%7 and 3 mm) are currently underway, which will help to find new
%component on the sub-pc scales and to clarify the identification
%of the components and the studies of polarisation in the jet,
%Faraday rotation, spectral index, else.

\begin{acknowledgements}

This research was supported by the National Natural Science
Foundation of PR China (19973103), and Chinese fund NKBRSF
(G1999075403). The authors thank Dr.Jim Ulvestad for his work on
the fringe-fitting and imaging the VSOP in-orbit checkout data of
1156+295. XYH thanks JIVE and ASTRON for their hospitality during
his visit to the Netherlands for the data reduction in 1999 and
2002. The studies reported in this paper have been supported by
the grant for collaborative research in radio astronomy of the
Dutch (KNAW) and Chinese Academies of Sciences (CAS). XYH thanks
Dr. Chris Carilli for the help of calibration for the polarization
VLBA data and Dr. Jun-Hui Zhao for the useful discussion on the
wobbling disk. The authors are grateful to the staff of the
European VLBI Network radio observatories and the data reduction
center, the MERLIN, the NRAO for support of the observations. The
European VLBI Network is a joint facility of European, Chinese,
South African and other radio astronomy institutes funded by their
national research councils. The National Radio Astronomy
Observatory is a facility of the National Science Foundation
operated under a cooperative agreement by Associated Universities,
Inc. The authors gratefully acknowledge the VSOP Project, led by
the Institute of Space and Astronautical Science in cooperation
with the world-wide network of agencies, institutes and
facilities. We thank the referee for critical reading of the
manuscript and many valuable comments. This research has made use
of data from the University of Michigan Radio Astronomy
Observatory which is supported by funds from the University of
Michigan.

\end{acknowledgements}


\begin{thebibliography}{}

\bibitem[Antonucci \& Ulvestad, 1985]{A1}
Antonucci, R. R. \& Ulvestad, J. S., 1985, ApJ, 294, 158

%\bibitem[Beck 2000]{Beck}
%Beck, R., 2000, in Perspecktive on Radio Astronomy, ed. M. P. van
%Haardlem (Dwingeloo: ASTRON), 249

%\bibitem[Beskin 1997]{Beskin}
%Beskin, V. S., 1997, Phys.-Uspeckhi, 40(7), 659

\bibitem[Blandford \& Rees 1974]{BR}
Blandford, R. D. \& Rees, M. J., 1974, MNRAS, 165, 395

\bibitem[Bower et al. 1997]{B1}
Bower, G. C., Backer, D. C., Wright, M., Forster, J. R., Aller, H.
D. \& Aller, M. F., 1997, ApJ, 484, 118

%\bibitem[Britzen et al 2000]{B2}
%Britzen, S., Witzel, A., Krichbaum, T. P., Campbell, R. M.,
%Wagner, S. J., \& Qian, S. J., 2000, A\&A, 360, 65
%
%\bibitem[Britzen et al 2000]{B3}
%Britzen, S., Roland, J., Laskar, J., Kokkotas, K., Campbell, R.
%M., \&  Witzel, A., 2001, A\&A, 374, 784
%
%\bibitem[Carilli \& Barthel, 1996]{C0}
%Carilli, C. L. \& Barthel, P. D.,1996, A\&A Annual Review, 7, 1-54

\bibitem[Conway  \&  Murphy,  1993]{C1}
Conway, J. \& Murphy, D. W., 1993, ApJ, 411, 89

\bibitem[1992]{CK92}
Camenzind, M. \& Krockenberger, M., 1992, A\&A, 255, 59.

%\bibitem[Fey et al. 1996]{F1}
%Fey, A. L., Clegg, A. W. \&  Fomalont, E. B., 1996, ApJS, 105, 299

\bibitem[Field \& Rogers 1993]{FR}
Field, G. B. \& Rogers, A. D., 1993, ApJ, 403 94

\bibitem[1999]{Fomalont} Fomalont, E. B., 1999, in Synthesis Imaging
in Radio Astronomy II, Taylor G. B., Carilli C.L., \& Perley R.A.
(eds.), P301

\bibitem[Fomalont et al. 2000]{F2}
Fomalont, E. B., Frey, S., Paragi, Z., Gurvits, L. I., Scott, W.
K., Taylor, A. R., Edwards, P. G., \& Hirabayashi, H., 2000, ApJS,
131, 95

\bibitem[Garrington et al. 1999]{STG1}
Garrington, S. T., Garrett, M. A., \& Polatidis, A., 1999, New
Astronomy Review, 43, 629.

\bibitem[Garrington et al. 2001]{STG2}
Garrington, S. T., Muxlow, T. \& Garrett, M. A., in Galaxies and
their constituents at the highest angular resolutions, IAU symp.,
205, Eds. Schilizzi, R.T., Vogel, S.N., Paresce, F., et al., P102.

\bibitem[Ghisellini et al. 1993]{Ghis93}
Ghisellini, G., Padovani, P., Celotti, A., \& Maraschi, L., 1993,
ApJ, 407, 65

\bibitem[Glassgold et al. 1983]{G1}
Glassgold, A. E., Bergman, J. M., Huggins, P. J., Kinney, A. L.,
Pica, A. J., Pollock, T. J., et al., 1983, ApJ, 274, 101

\bibitem[Gomez et al. 2001]{gomez}
Gomez, J.-L., Marscher, A. P., Alberdi, A., Jorstad, S. G., and
Agudo, I., 2001, ApJ, 561, L161

\bibitem[Guijosa \& Daly 1996]{G2}
Guijosa, A. \& Daly, R. A., 1996, ApJ, 461, 600

\bibitem[Gurvits et al. 2003]{G3}
Gurvits, L. I. Kellermann, K. I., Fomalon, E. B., Zhang, H. Y.,
2003, in preparation

\bibitem[Hardee 1987]{H1}
Hardee, P. E., 1987, ApJ, 318, 78

\bibitem[Hartman et al. 1999]{H2}
Hartman, R. C., Bertsch, D. L., Bloom, S. D., Chen, A. W.,
Deines-Jones, P., Esposito, J. A., et al. 1999, ApJS, 123, 79

%\bibitem[Hewitt \& Burbidge, 1989]{H3}
%Hewitt A. \& Burbidge G., 1989, ApJS, 69, 1

\bibitem[Hirabayashi et al. 1998]{H3}
Hirabayashi, H., Hirosawa, H., Kobayashi, H., Murata, Y., Edwards,
P. G., Fomalont, E. B. et al., 1998, Science, 281, 1825

%\bibitem[Hong et al. 1999]{H4}
%Hong, X. Y., Jiang, D. R., \& Shen, Z.-Q., 1998, A\&A Letter, 320,
%L45

\bibitem[Hong et al. 1999]{H5}
Hong, X. Y., Jiang, D. R., Gurvits, L. I., Garrett, M. A.,
Schilizzi, R. D., \& Nan, R. D., 1999, New Astronomy Review, 43,
699

\bibitem[Hong et al. 2003]{H6}
Hong, X. Y., et al.  2003, in preparation.

\bibitem[Jorstad et al. 2001]{J1}
Jorstad, S. G., Marscher, A. P., Mattox, J. R., Wehrle, A. E.,
Bloom, S. D., \& Yurchenko, A. V., 2001, ApJS, 134, 181

%\bibitem[Kellermann et al. 1998]{K1}
%Kellermann, K. I., Vermeulen, R. C., Zensus, J. A., \& Cohen, M.
%H., 1998, AJ, 115, 1295

\bibitem[Kellermann et al. 1999]{K2}
Kellermann, K. I., Vermeulen, R. C., Zensus, J. A., Cohen, M. H.,
\& West, A., 1999, New Astronomy Review, 43, 757


\bibitem[Kellermann \& Pauliny-Toth 1969]{KP}
Kellermann, K. I. \& Pauliny-Toth, I. I. K., 1969, ApJ, 155



\bibitem[Lehto \& Valtonen 1996]{LV}
Lehto, H. J., \& Valtonen, M. J., 1996, ApJ, 460, 270

\bibitem[Linfield, 1981]{L1}
Linfield, R. P., 1981, ApJ, 250, 464

\bibitem[Liu \& Melia 2002]{LM}
Liu, S. \& Melia, F., 2002, ApJ Lett, 566, L77

\bibitem[Lobanov 1998]{lobanov}
Lobanov, A. P. 1998, A\&A, 330, 79

\bibitem[Marscher 1987]{Mar87} Marscher, A.P. 1987, in Superluminal Radio Sources,
Zensus, J. A., \& Pearson, T. J. (eds.), Cambridge University
Press, New York, P208

\bibitem[McHardy 1985]{M0}
Mchardy, I., 1985, Space Sci. Rev. 40, 559

\bibitem[McHardy et al. 1990]{M1}
McHardy, I. M., Marscher, A. P., Gear, W. K., Muxlow, T., Lehto,
H. J., \& Araham, R. G., 1990, MNRAS, 246, 305

\bibitem[McHardy et al. 1993]{M2}
McHardy, I. M., Marscher, A. P., Gear, W. K., Muxlow, T., Lehto,
H. J., \& Araham, R. G., 1993, MNRAS, 261, 464

\bibitem[Mukherjee et al. 1997]{M3}
Mukherjee, R., Bertsch, D. L., Bloom, S. D., Dingus, B. L.,
Esposito, J. A., Fitchel, R.C., et al., 1997, ApJ, 490, 116

\bibitem[Piner \& Kingham, 1997]{P1}
Piner, B. G. \& Kingham, K., 1997, ApJ, 485, L61

\bibitem[Piner \& Kingham, 1998]{P1b}
Piner, B. G. \& Kingham, K., 1998, ApJ, 507, 706

%\bibitem[Pohl et al. 1995]{P2} Pohl, M., Reich, W., Krichbaum T. P.,
%Standke, K., Britzen, S., Reuter, H. P., et al. 1995, A\&A, 303,
%383

\bibitem[Readhead 1994]{Read94}Readhead, A. C. S., 1994, ApJ, 426, 51

\bibitem[Rees \& Simon, 1968]{R0} Rees, M. J. \& Simon, M., 1968,
ApJ, 152, 145

\bibitem[Roos, Kaastra \& Hummel 1993]{R1}
Roos, N., Kaastra, J. S. \& Hummel, C. A., 1993, ApJ, 409, 130

\bibitem[Scott \& Read 1977]{SR}Scott, M. A., \& Readhead, A. C. S., 1977, MNRAS, 180, 539

\bibitem[Shen et al. 1997]{Shen97}
Shen Z.-Q., Wan, T.-S., Moran, J. M., Jauncey, D. L., Reynolds, J.
E., Tzioumis, A. K.,  et al. 1997, AJ, 114, 1999

\bibitem[Shepherd et al. 1994]{S1}
Shepherd, M. C., Pearson, T. J., \& Taylor, G.B., 1994, BAAS, 26,
987

\bibitem[Sillanpaa et al. 1988]{Sill}
Sillanpaa, A., Haarala, S., Valtonen, M. J., et al. 1998, ApJ,
325, 628

\bibitem[Steffen et al 1995]{stef95} Steffen, W., Zensus, J. A.,
Krichbaum, T. P., Witzel, A., \& Qian, S. J., 1995, A\&A, 302, 335

\bibitem[Thomasson 1986]{THO}
Thomasson, P., 1986, QJRAS, 27, 413

\bibitem[Thompson et al. 1995]{TH}
Thomposon, D. J.,  Bertsch, D. L.,  Dingus, B. L., Esposito, J.
A.,  Ettenne, A., Fichtel, C. E., et al., 1995, ApJS, 101, 259.

\bibitem[Tornikoski \& Lahteenmaki 2000]{TL00}
Tornikoski, M. \& Lahteenmaki, A., Proceedings of the fifth
Compton Symposium, 2000. Edited by Mark L. McConnell and James M.
Ryan AIP Conference Proceedings, Vol. 510., p.377.

%\bibitem[Vermeulen \& Cohen, 1994]{V1}
%Vermeulen, R. C. \& Cohen, M. H., 1994, ApJ, 430, 467

\bibitem[V\'{e}ron-Cetty \& V\'{e}ron 1998]{VCV}
V\'{e}ron-Cetty, M.-P. \& V\'{e}ron, P., 1998, A Catalogue of
Quasars and Active Galactic Nuclei (8th edition), ESO Sci. Report
No. 18.

\bibitem[Villata et al. 1998]{villa}
Villata, M., Raiteri, C. M., Sillanpaa, S., \& Takalo, L. O.,
1998, MNRAS, 293, L13

\bibitem[Wehrle et al. 1998]{WW}
Wehrle, A. E., Pian, E., Urry, C. M., Maraschi, L., McHardy, I.
M., \& Lawson, A. J., 1998, ApJ, 497,178

\bibitem[Wills et al. 1983]{W1}
Wills, B. J., Wills, D., Breger, M., Antonucci, R. R., \&
Barvainis, R. E.,
 1983, ApJ, 274, 62

\bibitem[Wills et al. 1992]{W2}
Wills, B. J., Pollock, J. T., Aller, H. D., Aller, M. F., Balonek,
T. J., Barvainis, R. E., et al., 1992, ApJ, 398, 454

\bibitem[Zhao et al. 1992]{zhao}
Zhao, J.-H., Burns, J. O., Hardee, P. E., \& Norman, M. L., 1992,
ApJ, 387, 83

\bibitem[Zhou et al. 2000]{zhou}
Zhou, J. F., Hong, X. Y., Jiang, D. R., \&  Venturi, T., 2000, ApJ
Letter, 540, L13

%\bibitem[1986]{Zuckerman}
%Zuckerman, B., \& Dyck, H. M., 1986, ApJ, 311, 345


\end{thebibliography}
\end{document}